\documentclass[twocolumn]{aastex63}
\usepackage{natbib}
\usepackage{multirow}
\usepackage{amsmath}
\usepackage{xcolor}
\usepackage{CJK}
\graphicspath{{./}{figures/}}
\usepackage{color,subfigure,lineno} 

\newcommand{\hst}{{\it HST}}

\def\MgII{Mg\,{\sc ii}}
\def\FeII{Fe\,{\sc ii}}
\newcommand{\objfull}{SDSS~J082341.08$+$241805.0}
\newcommand{\obj}{SDSS~J0823$+$2418}

\defcitealias{Hwang20}{Paper~II}
\defcitealias{Shen19}{Paper~I}
\defcitealias{Chen22a}{Paper~III}
\defcitealias{Chen23}{Paper~IV}

\shorttitle{VODKA: \obj}
\shortauthors{Gross et al.}

\begin{document}
\begin{CJK*}{UTF8}{gbsn}

\title{Varstrometry for Off-nucleus and Dual sub-Kpc AGN (VODKA): Investigating the Nature of \obj\ at $z=1.81$: a Likely Lensed Quasar}

\correspondingauthor{Arran C. Gross}
\email{acgross@illinois.edu}

\author[0000-0001-7681-9213]{Arran C. Gross}
\affiliation{Department of Astronomy, University of Illinois at Urbana-Champaign, Urbana, IL 61801, USA}

\author[0000-0002-9932-1298]{Yu-Ching Chen}
\affiliation{Department of Astronomy, University of Illinois at Urbana-Champaign, Urbana, IL 61801, USA}

\author[0000-0002-1616-1701]{Adi Foord}
\affiliation{Kavli Institute of Particle Astrophysics and Cosmology, Stanford University, Stanford, CA 94305, USA}

\author[0000-0003-0049-5210]{Xin Liu}
\affiliation{Department of Astronomy, University of Illinois at Urbana-Champaign, Urbana, IL 61801, USA}
\affiliation{National Center for Supercomputing Applications, University of Illinois at Urbana-Champaign, Urbana, IL 61801, USA}

\author[0000-0003-1659-7035]{Yue Shen}
\affiliation{Department of Astronomy, University of Illinois at Urbana-Champaign, Urbana, IL 61801, USA}
\affiliation{National Center for Supercomputing Applications, University of Illinois at Urbana-Champaign, Urbana, IL 61801, USA}

\author[0000-0003-3484-399X]{Masamune Oguri}
\affiliation{Center for Frontier Science, Chiba University, Chiba 263-8522, Japan}
\affiliation{Department of Physics, Graduate School of Science, Chiba University, Chiba 263-8522, Japan}

\author[0000-0003-4700-663X]{Andy Goulding}
\affiliation{Department of Astrophysical Sciences, Princeton University, Princeton, NJ 08544, USA}



\author[0000-0003-4250-4437]{Hsiang-Chih Hwang}
\affiliation{School of Natural Sciences, Institute for Advanced Study, Princeton, 1 Einstein Drive, NJ 08540, USA}

\author[0000-0001-6100-6869]{Nadia L. Zakamska}
\affiliation{Department of Physics and Astronomy, Johns Hopkins University, Baltimore, MD 21218, USA}

\author[0000-0002-0463-9528]{Yilun Ma (马逸伦)}
\affiliation{Department of Astrophysical Sciences, Princeton University, Princeton, NJ 08544, USA}

\author[0000-0001-5769-0821]{Liam Nolan}
\affiliation{Department of Astronomy, University of Illinois at Urbana-Champaign, Urbana, IL 61801, USA}






\begin{abstract}
Dual quasars at small physical separations are an important precursor phase of galaxy mergers, ultimately leading to the coalescence of the two supermassive black holes. Starting from a sample of dual/lensed quasar candidates discovered using astrometric jitter in Gaia data, we present a pilot case study of one of the most promising yet puzzling candidate dual quasars at cosmic noon ($z\sim 1.8$). Using multi-wavelength imaging and spectroscopy from X-ray to radio, we test whether the \obj\ system is two individual quasars in a bound pair at separation$\sim$0.64\arcsec, or instead a single quasar being gravitationally lensed by a foreground galaxy. We find consistent flux ratios ($\sim$1.25$-$1.45) between the two sources in optical, NIR, UV, and radio, and thus similar spectral energy distributions, suggesting a strong lensing scenario. However, differences in the radio spectral index, as well as changing X-ray fluxes, hint at either a dual quasar with otherwise nearly identical properties, or perhaps lensing-based time lag of $\sim$3 days paired with intrinsic variability. We find with lens mass modeling that the relative NIR positions and magnitudes of the two quasars and a marginally detected central galaxy are consistent with strong lensing. Archival SDSS spectra likewise suggest a foreground absorber via \MgII\ absorption lines. We conclude that \obj\ is likely a lensed quasar, and therefore that the VODKA sample contains a population of these lensed systems (perhaps as high as 50\%) as well as dual quasars. 

\end{abstract}


\section{Introduction}\label{sec:intro}
Our current understanding of galaxy evolution indicates that the massive galaxies observed in the present-day universe have been built up in a series of major mergers of less massive galaxies \citep{DiMatteo05}. It is believed that nearly all massive galaxies contain a central supermassive black hole (SMBH) \citep{Kormendy95}, which grows primarily through accretion of mass \citep{YuTremaine02}, and whose evolution is linked to the growth of its host galaxy's stellar bulge \citep[yielding the $M_{\rm BH}-\sigma_{\star}$ relation,][]{McConnell13}. Galaxy mergers are often invoked to explain the growth necessary to produce observed massive SMBHs \citep[$e.g.,$][]{Zhang21}. Such phases of accretion via gas inflows to the SMBH also trigger periods of immense activity \citep[e.g.,][]{Hernquist95, VanWassenhove12, Capelo16}, seen observationally as an active galactic nucleus (AGN), the most extreme of which are classified as quasars ($L_{\rm bol}\gtrsim10^{45}$erg s$^{-1}$). 

It is predicted that at small physical separations ($\lesssim$10 kpc) in the later stages of a galactic merger \citep[e.g,][]{VanWassenhove12, RosasGuevara18}, both SMBHs in the constituent galaxies have a significant chance of being active simultaneously as dual quasars. The pair will eventually form a more compact binary SMBH, ending in a BH merger event \citep{Begelman80}. Identifying dual quasar systems at small separations is thus essential for constraining the rate of binary black hole mergers, which in turn is needed to inform predictions of gravitational waves emitted as a consequence of those mergers \citep[e.g.,][]{Peters64, Dotti12, Abbott16, Holgado19, Goulding19}. The activity of galaxy mergers and quasars is thought to peak around the epoch of cosmic noon ($z\sim2$) \citep{Richards06}. However, despite decades of searches, both serendipitous and systematic discoveries of dual quasars have remained fairly rare \citep[see, e.g.,][and references therein]{Satyapal17, DeRossa20, Gross23a}, yielding even fewer confirmed dual AGN at the high redshift and small separations that are most relevant for probing cosmic growth \citep[see][for a recent review]{Chen22a}. The lack of observed dual quasars is partially due to their apparent intrinsic rarity. Using Gaia EDR3-resolved pairs, \citet{Shen23} estimate that the fraction of double quasars (3 kpc$<$sep$<$30 kpc) among luminous unobscured quasars at 1.5$<z<$3.5 is $\sim6.2\pm0.5\times 10^{-4}$. Limitations of current observatories to probe such small spatial scales in wide area surveys also hamper efforts to identify promising candidates. 

The confirmation of dual quasar candidates then requires multi-wavelength follow-up observations, where systematic issues can lead to spurious classifications. For example, X-ray observations can unambiguously classify bright sources as AGN, but the low spatial resolution of X-ray telescopes ($\gtrsim$0.5\arcsec\ for on-axis observations with $Chandra$) precludes identifying dual sources with small separations. While IR surveys using the Wide-field Infrared Survey Explorer \citep[WISE;][]{Wright10} have had much success in systematically recovering the obscured AGN population \citep{Barrows21}, resolution limitations also hamper robust detection of dual quasars with separations on the kiloparsec scale (e.g., the $W2$ band has a point spread function (PSF) full width at half maximum (FWHM) of 6\arcsec, corresponding to a separation of $\sim$50 kpc at $z=2$). 

A contaminant population for the dual quasar search is that of small-separation gravitationally lensed single quasars. At high redshift, lensing galaxies can be difficult to detect due to limited spatial resolution, as well as their inherent faintness compared to the more luminous and potentially blended quasar images. Indeed, many systematic searches for dual quasars have turned up lensed quasars incidentally \citep[see][for a recent review]{Shajib22}. These systems are intriguing in their own right; cases of strong gravitational lensing have been used in recent decades to address cosmological questions including the value of the Hubble constant \citep{Kochanek06}. In particular, lensed quasars are novel probes of dark matter, and high-resolution imaging of lensed quasar radio jets have been used to constrain the so-called ``fuzzy" dark matter model \citep[][and references therein]{Hui17, Powell23}. Accurately determining the nature of a given dual quasar candidate as a bona fide dual AGN or a single lensed quasar is therefore critically important to its astrophysical implications.  

Throughout this series of papers, we have endeavored to probe the hitherto poorly explored regime of high-$z$ dual quasar candidates at sub-arcsec separations through Varstrometry for Off-nucleus and Dual sub-Kpc AGN (VODKA). The technique of varstrometry was first described by \citet{Shen19} and \citet{Hwang20}, hereafter \citetalias{Shen19} and \citetalias{Hwang20}, respectively. Briefly, varstrometry relies on variable jitters in the astrometric centroid position of an unresolved source over a series of observations. Such a signature can indicate the presence of two separate sources with intrinsic photometric variability, making it possible to compile samples of candidate dual quasars at kiloparsec-scale physical separations.

In \citetalias{Hwang20} we compiled a sample of candidate dual quasars by primarily applying the varstrometry selection to the Gaia Data Release 2 (DR2) catalog \citep{Gaia18}, as well as searching for resolved pairs. Using a series of selection criteria based on astrometric noise and cross-matches to SDSS, WISE, and Pan-STARRS, a catalog of 150 targets was assembled in \citet{Chen22a}, hereafter \citetalias{Chen22a}. From this catalog 84 targets were subsequently observed in \citetalias{Chen22a} as part of follow-up two-band Hubble Space Telescope ($HST$) imaging (SNAP-15900, PI: Hwang) to obtain optical photometric colors, yielding a sample of VODKA targets composed of 45 resolved pairs in Gaia with separations 0.4\arcsec$<$sep$<$0.7\arcsec\ \citep[limited by the minimal separation in Gaia DR2;][]{Arenou18}. Recently, we conducted a pilot study in \citet{Chen23}, hereafter \citetalias{Chen23}, focused on one of these dual quasar candidates, SDSS J0749+2255, first reported in \citet{Shen21}. A robust combination of multi-wavelength imaging and spectroscopy confirmed this system to be a dual quasar. The system is composed of two $\sim10^{9}M_{\odot}$ SMBHs hosted by a disk-disk galaxy merger with tidal features and stellar masses of $\sim10^{11.5}M_{\odot}$, making it the first discovery of such a system at nuclear separation of $\lesssim$4 kpc at cosmic noon. 

\begin{figure}
     \centering
         \includegraphics[width=0.46\textwidth]{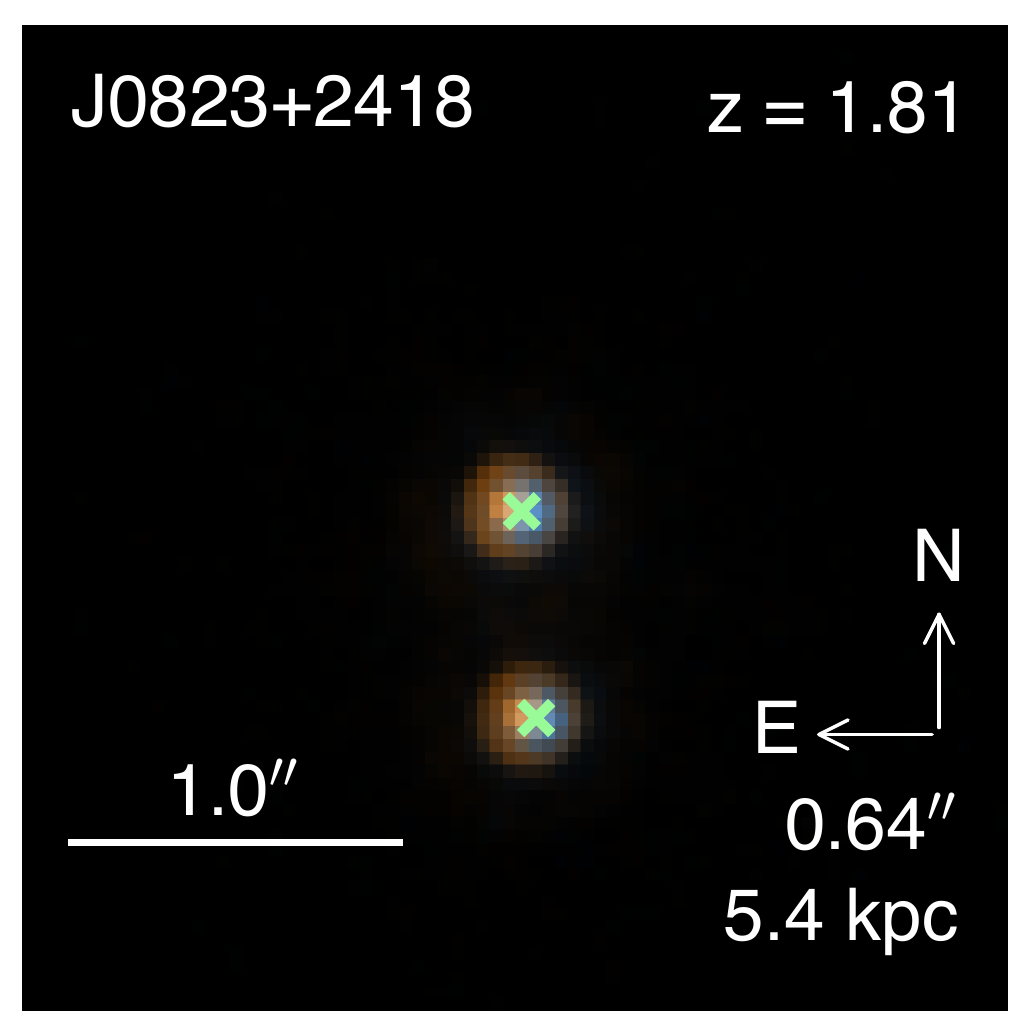}
        \caption{HST/WFC3 color composite image from \citetalias{Chen22a}, where F475W is coded in blue, F814W in red, and the average of F475W and F814W bands in green). The Gaia DR2 source positions are marked by the green crosses. The angular and physical separations between the double sources are given in the bottom right corner.
}
     \label{fig:HST_img}
\end{figure}

In this paper, we build on the previous work by performing a case study on another promising VODKA target, \objfull, hereafter \obj. Our main goal in this work is to determine whether the \obj\ system is a dual quasar or a lensed single quasar, and to characterize its properties. \obj\ is spatially resolved by Gaia, and received preliminary follow-up imaging in \citetalias{Chen22a} using $HST$ WFC3 bands F475W and F814W, shown in Figure \ref{fig:HST_img}. The difference in photometric colors suggest that both the North (N) and South (S) components are quasars, and not foreground stars (see \citetalias{Chen22a} for selection criteria). We also consider \obj\ a promising candidate dual quasar because it exhibits two distinct detections in preliminary radio imaging, indicative of two radio AGN with different spectral indices. If confirmed as a dual quasar, \obj\ would be part of a select population of high redshift ($z\sim1.81$) systems at an advanced merger stage (sep$\sim5.4$ kpc).   

We organize the paper as follows. In \S\ref{sec:obs} we detail the multi-wavelength observations and their associated data reduction procedures. In \S\ref{sec:result} we perform various analyses and give the general interpretation case by case. We combine all of the results as lines of evidence in \S\ref{sec:discuss} to weave together a coherent picture of \obj. We also compare it to similar systems from the literature. In \S\ref{sec:sum} we summarize our findings and offer concluding remarks for future campaigns with similar strategies. Throughout this work, we assume a flat $\Lambda$CDM cosmology with values of $\Omega_{\rm \Lambda}$=0.7, $\Omega_{\rm m}$=0.3, and $H_{0}$=70 km s$^{-1}$ Mpc$^{-1}$.

\section{Observations and Analysis}\label{sec:obs}
To assess whether the \obj\ system is a dual quasar, we employ multi-wavelength observations ranging from X-rays to radio, using a mix of photometry and spectroscopic observations. In some wavebands, we have obtained complementary observations to account for various differences in observational performance, {\it i.e.}, higher angular resolution versus deeper observations for faint structure.  

In addition to the observations detailed below, we have also acquired slit spectroscopy observations using the Magellan Folded-port Infrared Echellette (FIRE) spectrograph. However, the redshift of \obj\ places the prominent rest-frame optical emission lines (H$\alpha$, [O {\sc iii}], and H$\beta$) in windows that are heavily affected by telluric absorption, even after correction. The traces of the 2D spectra are only marginally separated, complicating the extraction. While the broad H$\gamma$ emission line does show some differences between the N and S sources, there might be some contamination from Fe {\sc ii} emission. We do not discuss this observation further owing to these complications. 
\subsection{HST Imaging}\label{sec:hst}
Infrared (IR) imaging in the observed frame has the potential to uncover low-surface brightness features critical for the interpretation of the \obj\ system. For example, detection of faint tidal features, such as trailing galactic arms, would be highly suggestive of an ongoing galactic merger between the two nuclei. Conversely, detection of a faint central galaxy between the two nuclei could be evidence of gravitational lensing. Both of these scenarios require careful subtraction of the bright point source quasars to uncover the fainter residual features. 

We obtained high-resolution imaging of \obj\ using the \hst\ Wide Field Camera 3 (WFC3) on 8 February 2022 (Program GO-16892; PI: X. Liu). We use the NIR range F160W filter to capture the faint underlying hosts. This filter operates in the wide (width 168.3 nm) $H$-band ($\lambda_{\rm pivot}$ = 1536.9 nm), corresponding to a wavelength of $\lambda\sim546$ nm in the rest-frame of the host galaxy. We obtain a net exposure time of 2055 s. The detector's PSF (FWHM = 0.15\arcsec) is undersampled by the pixel scale (0.13\arcsec) necessitating dithering during the observations. The dithered frames are stacked during data reduction following standard routines for WFC3 \citep{Sahu21}, which clean the frames for cosmic ray trails and pixel area effects. We show the processed F160W image in the leftmost panel of Figure \ref{fig:HST_galfit}. The reduced image has a pixel scale of 0.06\arcsec\ and a photometric zeropoint of 25.936 (28.177) in the AB(ST) system. We reach a surface brightness limit of $\sim$25.65 mag/arcsec$^{2}$.

\subsection{Keck Imaging}\label{sec:keck}
We pursue a similar approach as with $HST$ using NIR imaging from the Keck Telescope under Program N072 (PI: Y. Shen). The adaptive optics (AO)-assisted NIRC2 camera is able to achieve a finer resolution than $HST$, with a PSF FWHM of 0.05\arcsec\ sampled at 0.01\arcsec/pix. This enables us to better model the point sources and extended host components so that a central lens galaxy can be more accurately constrained. 

Observations of \obj\ were conducted using the Keck NIRC2 camera on 18 December 2021 UT. We observed the target using the NIR $K_{p}$ filter, which has a central wavelength of 2.124 $\mu$m and width of 0.351 $\mu$m. The nearby star N910000396, 41.68\arcsec\ from \obj, was used as a tip-tilt reference star. Dithering was done using a 4\arcsec$\times$4\arcsec\ 9-position box. After standard reduction procedures, we co-add the exposures for a total exposure time of 1260 s. The Laser Star Guide mode was used for the adaptive optics corrections for atmospheric distortions during observations, and we subsequently use the guide stars to compute the photometric  zeropoint of the $K_{p}$ filter to be 26.56 (AB). We reach a surface brightness limit of $\sim$24.94 mag/arcsec$^{2}$. We show the reduced $K_{p}$ image in the leftmost panel of Figure \ref{fig:Keck_galfit}.

\subsection{HST/STIS Spectroscopy}\label{sec:stis}
            Observations of \obj\ using the $HST$/Space Telescope Imaging  Spectrograph (STIS) were obtained on 13 January 2022 (Program GO-16210; PI: X. Liu).  A 0.2\arcsec\ slit was used for a net exposure time of 897 s. at a position angle of 48.6$^{\circ}$ E of N to cover both nuclei simultaneously. The 5030\AA\ bandwidth ranges from 5236\AA\ to 10266\AA, with a spectral resolution of $\sim$790 at 7751\AA. The spatial resolution of PSF FWHM$\sim$0.08\arcsec\ allows for a reasonable separation of the two nuclei. 

\begin{figure}
     \centering
         \includegraphics[width=0.48\textwidth]{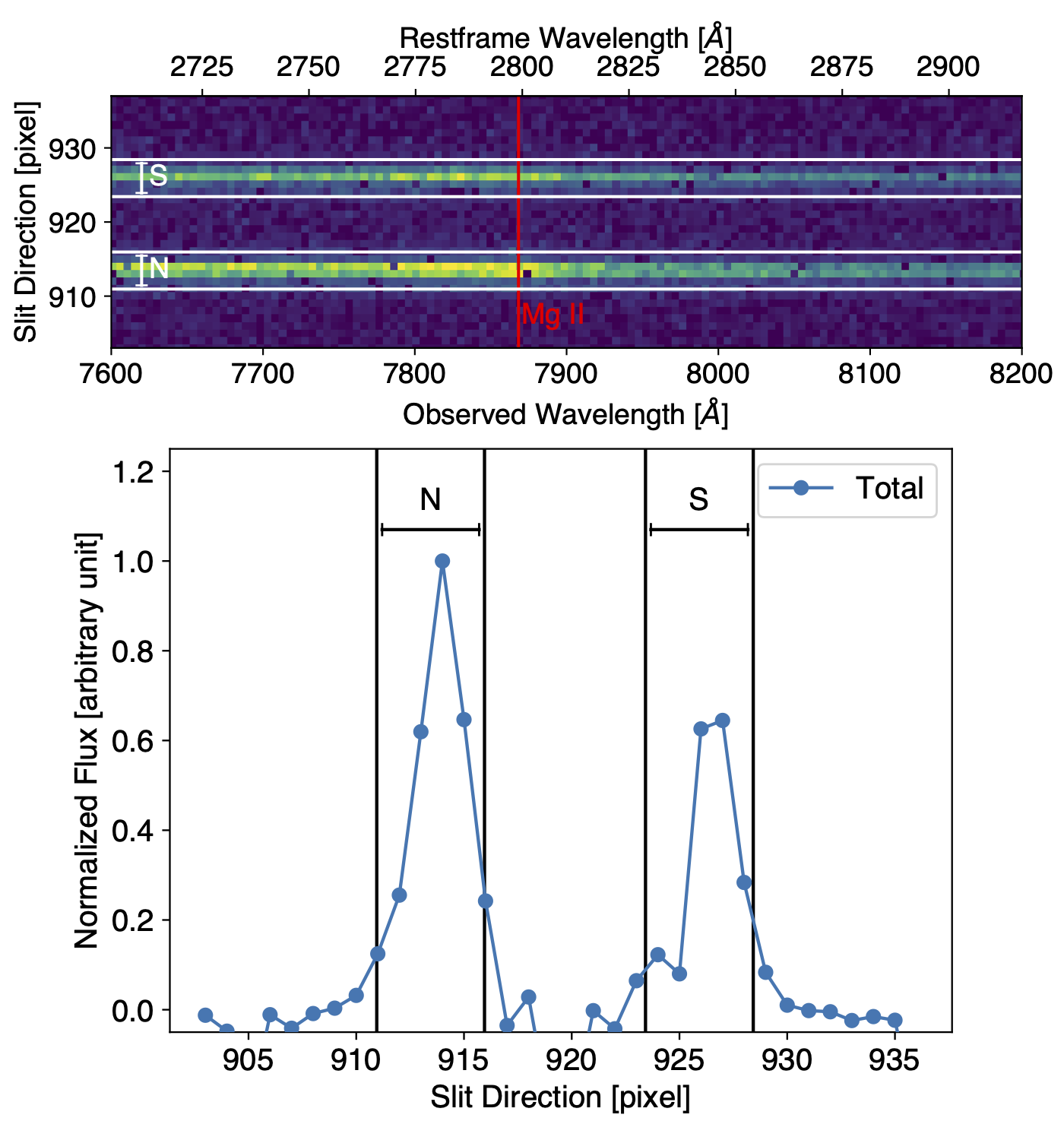}
        \caption{STIS spectral extraction. {\bf Top:} 2D spectrum with the extraction regions marked in white. We highlight the position of the bright \MgII\ emission line. {\bf Bottom:} Collapsed 1D profiles and corresponding extraction regions. }
     \label{fig:STISredux}
\end{figure}

\begin{figure*}
     \centering
         \includegraphics[width=\textwidth]{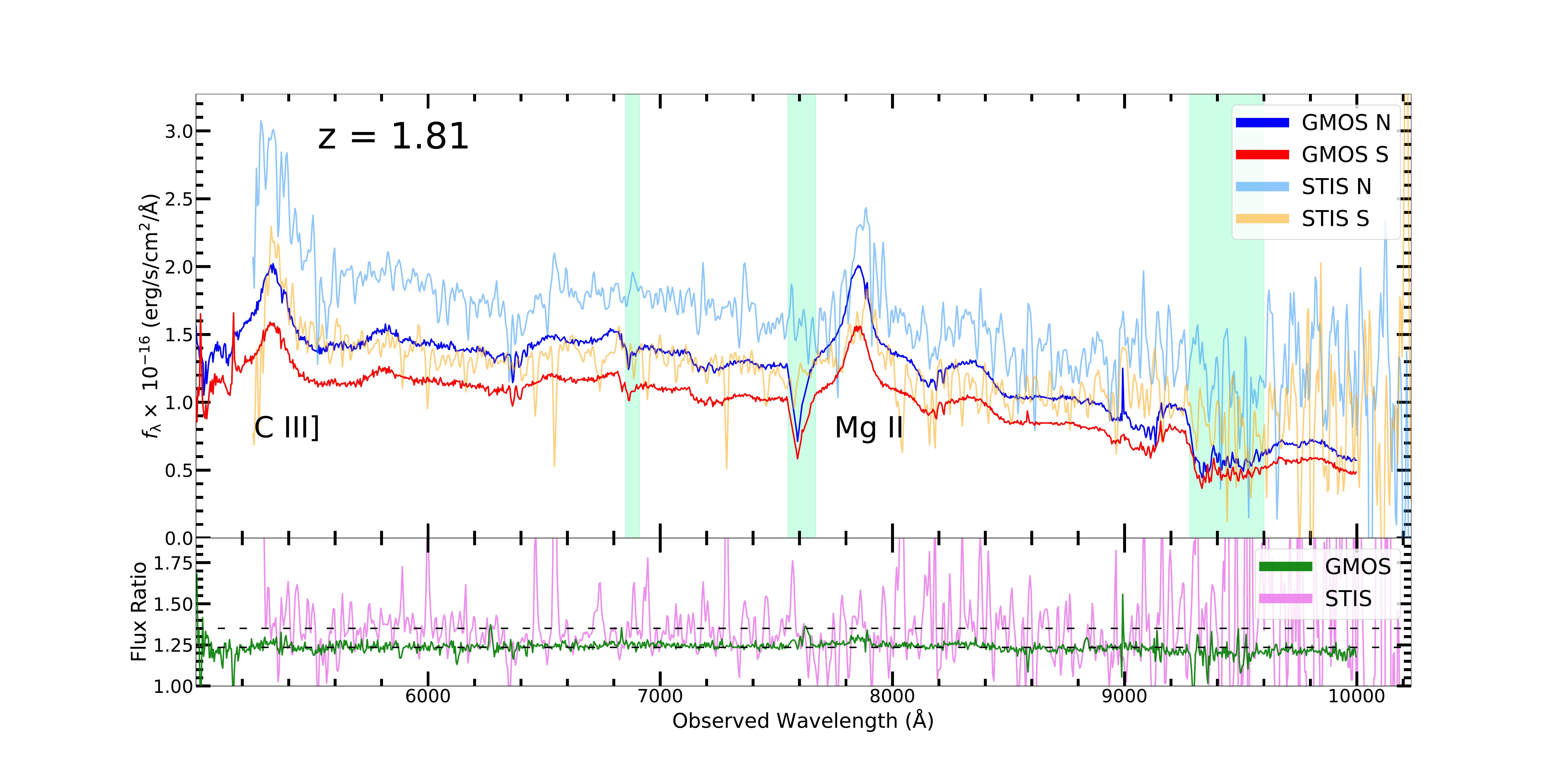}
        \caption{Observed-frame optical spectra for the two quasars. {\bf Top:} We show the HST/STIS spectra for \obj\ as well as the Gemini/GMOS spectra, indicated by the legend. The STIS spectra have been smoothed with a Gaussian kernel with sigma = 1 for readability. Green shaded regions are heavily affected in the GMOS spectra by telluric absorption. {\bf Bottom:} Flux ratios between the N and S quasars for the STIS and GMOS spectra. Dashed lines show the average of the corresponding flux ratio, with values that are roughly consistent with those found with photometry. The N and S spectra from both instruments are quite similar, suggestive that the system is a single lensed quasar where the N source is more magnified than the S component.}
     \label{fig:opspec}
\end{figure*}

We reduce the spectra using the \texttt{stis\_cti} reduction package \citep{Anderson10}, accounting for flat-fielding, combining exposures, and removing trails from cosmic rays and other artefacts. In Figure \ref{fig:STISredux}, we show the 2D spectrum highlighting the separation of the traces for the two nuclei, as well as the collapsed 1D brightness profiles and the extraction apertures. At an angular separation of 0.64\arcsec, we are able to obtain reasonable extractions for the two nuclei using apertures with widths of 5 pixels, corresponding to 0.25\arcsec, shown in Figure \ref{fig:STISredux}. After wavelength and flux calibration, we obtain the final 1D spectra for each nucleus, shown in Figure \ref{fig:opspec}.

\subsection{Gemini/GMOS Spectroscopy}\label{sec:gemini}
As an independent verification of the STIS spectra, we obtained optical/NIR spectra for \obj\ using the Gemini Multi-Object Spectrograph (GMOS) on the Gemini North Telescope on the night of 20 April, 2022 (Program ID GN-2022A-Q-139; PI: X. Liu). GMOS has wavelength coverage that goes further into the blue (5000\AA-10000\AA) than STIS, allowing us to fully cover the C{\sc iii}] quasar emission line (rest-wavelength $\lambda$1909\AA). We used the R150 grating and 0.75\arcsec slit oriented at a position angle of 3.2$^{\circ}$ E of N to cover both nuclei. The spectral resolution of GMOS in this configuration ($R\sim$631) is slightly lower than that of the STIS observation above, as is the spatial resolution with PSF 0.39\arcsec$<$FWHM$<$0.49\arcsec. However, the deeper observations allow for higher signal to noise compared to the STIS data.  

\begin{figure}
     \centering
         \centering
         \includegraphics[width=0.49\textwidth]{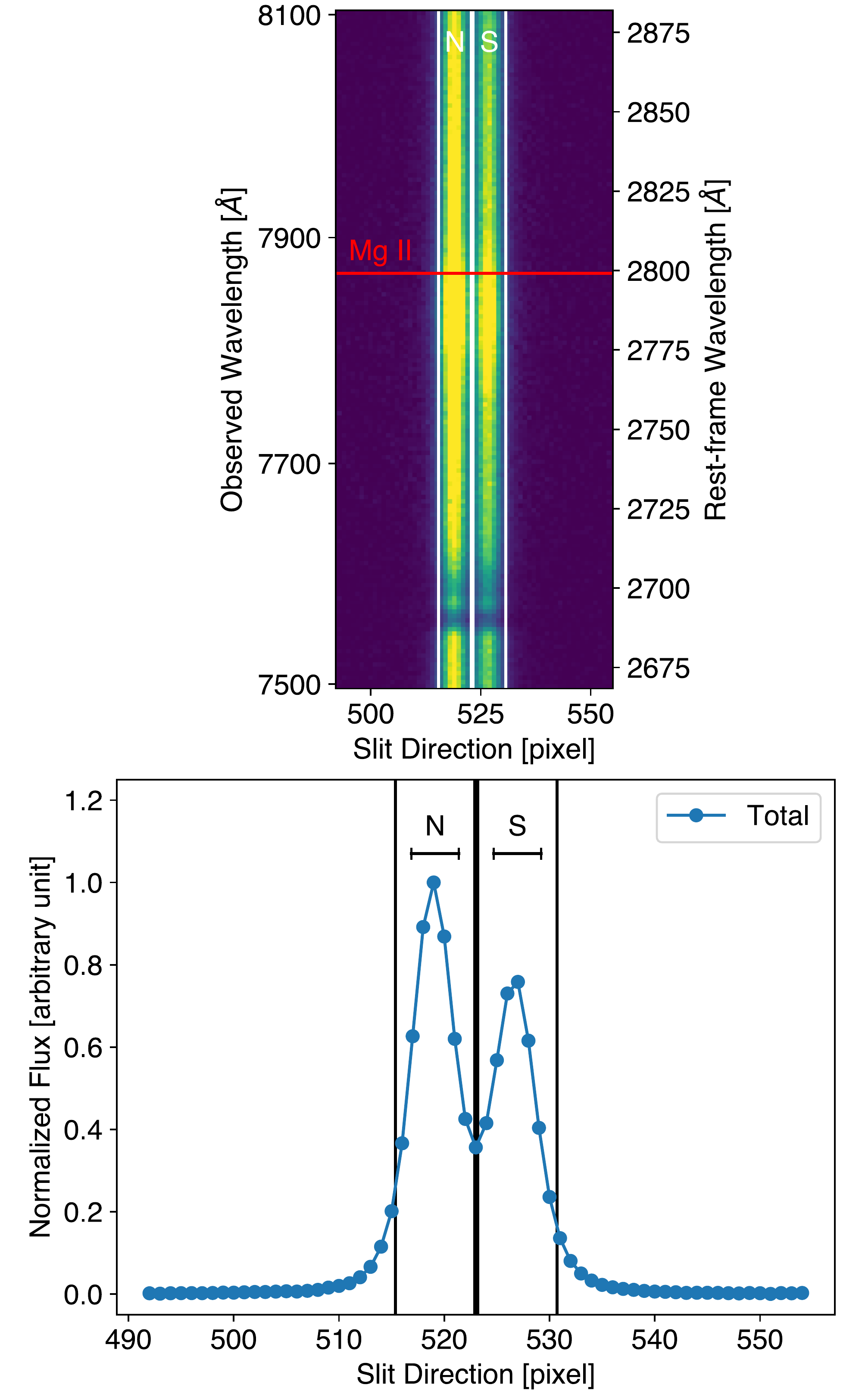}
        \caption{GMOS spectral extraction. {\bf Top:} 2D spectrum with the extraction regions marked in white. We highlight the position of the bright \MgII\ emission line. {\bf Bottom:} Collapsed 1D profiles and corresponding extraction regions. }
     \label{fig:GMOSredux}
\end{figure}


We reduce the raw spectrum using the \texttt{pypeit} spectral processing package \citep{Prochaska20a, Prochaska20b}, set to \texttt{gemini\_gmos\_north\_ham} mode. The pipeline handles the typical reduction steps correcting for flat-fields, bias, and cosmic rays. We manually set apertures with widths of 7.43 pixels (corresponding to 0.6\arcsec) for boxcar extraction of the 1D spectrum for each nucleus. In Figure \ref{fig:GMOSredux}, we show the 2D spectrum, as well as the collapsed 1D brightness profiles and the extraction apertures. Extraction is less straightforward with the GMOS spectrum than the STIS spectrum above due to the larger FWHM of the PSF. Thus, there is some degree of source blending in the extraction apertures. Wavelength calibration is done using the CuAr spectra acquired during observations. Flux calibration is done using observations of the standard star EG 131. We do not use \texttt{pypeit} telluric corrections as they are not yet finalized in the pipeline; we thus mask heavily affected regions in our subsequent treatment. We show the two calibrated spectra from GMOS in Figure \ref{fig:opspec}.       

\subsection{VLA Imaging}\label{sec:vla}
\begin{deluxetable}{lccccr}
\tabletypesize{\small} 
\tabletypesize{\scriptsize}

\tablecaption{VLA Observations
\label{tab:VLA_obs}}
\tablehead{ 
\colhead{Frequency} & \colhead{UT} & \colhead{IT} & \colhead{rms} & \colhead{Beam} & \colhead{PA} \\
\colhead{Band}  & \colhead{date} & \colhead{min} & \colhead{$\mu$Jy/beam} & \colhead{(arcsec)} & \colhead{(deg)} \\
\colhead{(1)} & \colhead{(2)} & \colhead{(3)} & \colhead{(4)} & \colhead{(5)} & \colhead{(6)} 
}
\startdata 
C (6 GHz) & 2020-12-13 & 30 & 8.4189 & 0.92$\times$0.35 & $-$69 \\
Ku (15 GHz) & 2020-12-10 & 30 & 7.3683 & 0.11$\times$0.10 & $+$18\\
\enddata
\tablecomments{ 
(1) Frequency band and nominal central frequency;
(2) UT date of observation (Y-M-D);
(3) Integration time for the observation, encompassing the kinematic pair;
(4) rms noise for the cleaned image;
(5) Restoring beam size (maj$\times$min axes);
(6) Restoring beam position angle.
}
\end{deluxetable}

Imaging \obj\ at radio frequencies has several advantages that make it complimentary to optical and IR observations. The spectral slope of the radio emission can be used as a tracer of synchrotron emission from jet activity, and differences between the spectral index values of the two sources could indicate a system of two distinct quasars instead of a single lensed source. 

We obtained radio imaging of \obj\ using the NSF's Karl G. Jansky Very Large Array (VLA) in two frequency bands: C-band (central frequency 5.9985 GHz, width of $\sim$4 GHz) and Ku-band (central frequency 14.9984 GHz, width of 6 GHz). Observations were conducted as part of Program 20B-242 (PI: X. Liu), and both bands were observed in the VLA's A-configuration. The source 3C147 was also observed and used for flux and bandpass calibration via the standard VLA calibration and flagging pipeline. We use the Common Astronomical Software Applications (CASA) package version 5.6.1 for data reduction and analysis of the Stokes {\it I} images. We reduce the images using the synthesis imaging and deconvolution routine \texttt{tclean} in w-project mode to account for sky curvature. We use Briggs weighting with robustness of 0.5 to achieve a balance between resolving small scale structure (i.e., separating the two nuclei) while also preserving low surface brightness features. The pixel scales are set to 0.1\arcsec\ and 0.03\arcsec\ for the C-band and Ku-band images, respectively, so that 3$-$5 pixels are sampled by the restoring beams. We set cleaning to run until either 1000 iterations are completed, or the rms residuals drop below 12 $\mu$Jy beam$^{-1}$. The properties of the restoring beams and noise levels for each image are given in Table \ref{tab:VLA_obs}. In the leftmost panels of Figure \ref{fig:VLA} we show the cleaned radio continuum images at 6 GHz and 15 GHz.

\subsection{Chandra Imaging and Spectroscopy}\label{sec:chandra}
Observations of \obj\ using the {\it Chandra X-ray Observatory} Advanced CCD imaging spectrometer \citep[ACIS;][]{Garmire03} -S  array were split into two observation blocks under Program 23700377 (PI: X. Liu; ObsIDs 25385 and 26279, hereafter referred to as observations 1 and 2, respectively). These correspond to MJDs 21 and 24 of January 2022. Both observations were conducted with the target close to the aim-point on the S3 chip, for durations of 3.83 and 14.86 ks, respectively. We reduce and analyze the observations using the {\it Chandra} Interactive Analysis of Observations software ({\sc ciao}) Version 4.14, with calibrations from {\sc caldb} 4.9.8. 

The datasets are reprocessed using \texttt{chandra\_repro} with the energy-dependent sub-pixel event repositioning algorithm \citep[EDSER;][]{Li04}, subsampling the pixels to 1/2 the native size. We then perform an astrometric correction by detecting sources in the level-2 event file via \texttt{wavdetect}, which we then cross-match to optical sources from the Sloan Digital Sky Survey Data Release 9 (SDSS DR9) catalog. For each observation, we require at least 3 matches to within a tolerance of 2\arcsec. We find that background flaring is within 3$\sigma$ of the mean background level and is thus negligible for all time intervals. In the top row of Figure \ref{fig:chandra}, we show the reduced data for the two observations in the broad 0.5$-$7 keV band.

\section{Results}\label{sec:result}
Differences in fluxes at a given wavelength between the N and S quasars would be suggestive of two unique sources constituting a dual quasar system. However, if the flux ratio is constant as a function of wavelength, then the  difference might be due to wavelength-independent lensing magnification instead. In this section we test the imaging and spectroscopic data for such differences. From \citetalias{Chen22a}, we have $HST$ photometric magnitudes for two optical bands, which we list in Table \ref{tab:phot}.

\begin{deluxetable}{cccccccc}
\tabletypesize{\small} \tablewidth{9pt}
\tabletypesize{\scriptsize}
\tablewidth{\textwidth}

\tablecaption{General Properties of \obj\
\label{tab:phot}}
\tablehead{
\colhead{Target} & \colhead{Redshift} & \colhead{F475W} & \colhead{F814W}  & \colhead{log${\it M}_{\rm BH, vir}$}\\
\colhead{} & \colhead{} & \colhead{mag} & \colhead{mag}  &   \colhead{${\it M}_{\odot}$} \\
\colhead{(1)}& \colhead{(2)}& \colhead{(3)}& \colhead{(4)}& \colhead{(5)}
}
\startdata 
N & 1.81354$\pm0.00036$ & 18.49$\pm$0.04 & 17.71$\pm$0.04  & 9.53$\pm$0.50 \\
S & 1.81328$\pm0.00036$ & 18.88$\pm$0.04 & 18.03$\pm$0.04  &  9.55$\pm$0.54 
\enddata
\tablecomments{ 
(1) Target \obj; (2) redshift measured from best-fit STIS \MgII\ emission line center; (3)-(4) archival photometric magnitudes (AB) from \hst\ \citep{Chen22a}; (5) Black hole mass derived from virial mass estimates using \MgII.  
}
\end{deluxetable}

\subsection{IR Image Decomposition}\label{sec:galfit}

To model the surface brightness profiles of each point source quasar, we first generate a PSF model. The $HST$ PSF is known to vary during oribits due to ``breathing" \citep[$e.g.,$ ][]{Lallo05}. Another complication is that our F160W observation of \obj\ does not have any field stars in frame to use as the PSF basis. We therefore construct an average PSF using 8 field stars from 6 other VODKA observations that were taken within $\sim$1 month of \obj\ and use this as our baseline PSF model.

\begin{figure*}
     \centering
         \includegraphics[width=\textwidth]{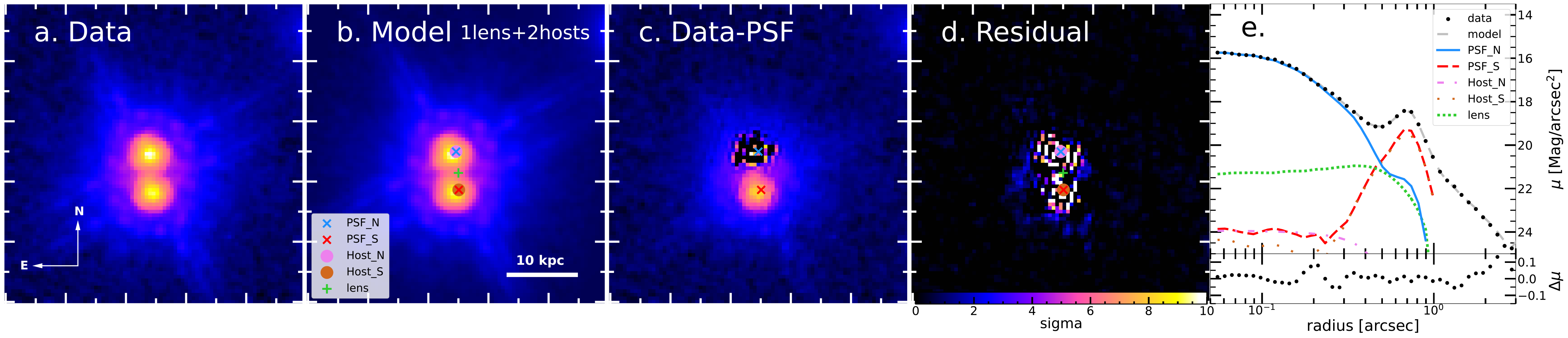}
        \caption{$HST$ F160W imaging spatial decomposition with GALFIT. {\bf a:} imaging data. {\bf b:} Model 3, where the legend gives the two quasar point sources and lens galaxy S\'ersic profile. {\bf c:} the image with only the point sources subtracted, leaving residuals due to the lensing galaxy but also the strong contribution from the S host galaxy. {\bf d:} the full model subtracted from the data, where the residuals are scaled by $\sigma$ significance above the remaining background. While the PSF models leave residuals corresponding to the host galaxies, the central lensing galaxy is well constrained. {\bf e:} radial profile starting from the center of the N PSF. The lower sub-panel shows the scaled residuals as a function of radius. Model 3 does a reasonably good job out to a radius of $\sim$1\arcsec, beyond which the brightness from the extended host galaxies falls off rapidly. The S host component closely follows the S PSF component, while the N host component is heavily suppressed. Major tick marks in panels a-d are 0.5\arcsec.}
     \label{fig:HST_galfit}
\end{figure*}

We use the software {\sc galfit} \citep{Peng10} to perform a 2D spatial decomposition of the NIR image from $HST$. Our zeroth-order model, given as Model 0 in Table \ref{tab:HST_gal}, includes the two PSF components for the N and S sources where the locations are allowed to vary, as well as a constant background. We also include components for two sources that are $>$7\arcsec\ to the NW of the N quasar (just outside of the frame in Figure \ref{fig:HST_galfit}). The inclusion of these other galaxies does not affect the best-fit parameters of the quasars, so we do not discuss them further; however, their inclusion improves the overall fit statistics, so they are included in all of our models going forward. These two external sources are modeled with a simple S\'ersic profile:
\begin{equation}
    \Sigma(r) = \Sigma_{\rm e}{\rm exp}\left[ -\kappa\left( \left(\frac{r}{r_{\rm e}}\right)^{1/{\rm n}}-1\right) \right],
\end{equation}
where $\Sigma(r)$ is the pixel surface brightness as a function of radial distance $r$ from the center, $\Sigma(r_{\rm e})$ is the pixel surface brightness at the ``effective" half-light radius $r_{\rm e}$, $n$ is the S\'ersic index, and $\kappa$ is a parameter related to n via complete and incomplete Gamma functions. The radius $r$ also encapsulates the ellipticity and position angle of the profile.

The statistical uncertainties on the best-fit parameters are known to be underestimated by {\sc galfit} \citep[$e.g.,$][]{Haussler07}, and the variations in the $HST$ PSF likely contribute large systematic uncertainties. To more accurately assess the PSF contribution, we perform the model fitting again using each of the 8 individual field star PSF models. For each fitted parameter, we then take the standard deviation across the 8 individual runs and baseline model, and add this to the statistical uncertainties output by {\sc galfit} to quote the total uncertainties in Table \ref{tab:HST_gal}.

Model 0 fails to capture the diffuse emission that extends $\sim$1.5\arcsec\ out from the center of the system. The residuals appear to be particularly concentrated between the two point sources instead of showing a diffuse extended morphology. The location of the residuals is suggestive of a faint central galaxy between the two quasars. Under the supposition that the two quasar targets might actually be a single quasar strongly lensed by a central foreground galaxy located between the N and S source locations, we define a more complex Model 1 which includes all of the components of Model 0 plus the lens galaxy component. The brightness distribution of this lens galaxy component is modeled using a S\'ersic profile.  

\begin{deluxetable}{lcc}
\tabletypesize{\small} \tablewidth{9pt}
\tabletypesize{\scriptsize}
\tablewidth{\textwidth}

\tablecaption{HST (F160W) Spatial Modeling
\label{tab:HST_gal}}
\tablehead{ 
\colhead{Parameter} & \colhead{\obj\ N} & \colhead{\obj\ S}  
}
\startdata 
\multicolumn{3}{c}{Model 0 (2 PSFs): $\chi^{2}_{\nu}$ =  7.777}  \\ \hline
m$_{\rm PSF}$ (AB) & 17.929$\pm$0.002 & 18.209$\pm$0.002 \\
C$_{\rm PSF}$ & 08:23:41.09+24:18:05.5 & 08:23:41.09+24:18:04.9 \\
\hline
\multicolumn{3}{c}{Model 1 (2 PSFs + 1 S\'ersic (lens)): $\chi^{2}_{\nu}$ =  4.357}  \\ \hline
m$_{\rm PSF}$ (AB) & 17.960$\pm$0.001 & 18.297$\pm$0.001 \\
C$_{\rm PSF}$ & 08:23:41.09+24:18:05.5 & 08:23:41.09+24:18:04.9 \\
m$_{\rm lens}$ (AB) &  \multicolumn{2}{c}{ 19.522$\pm$0.009 } \\
C$_{\rm lens}$ &  \multicolumn{2}{c}{ 08:23:41.09+24:18:05.1 } \\
$R_{\rm lens}$ (kpc) &  \multicolumn{2}{c}{3.27$\pm$0.05 } \\
n$_{\rm lens}$  & \multicolumn{2}{c}{1.93$\pm$0.05 }\\
\hline
\multicolumn{3}{c}{Model 2 (2 PSFs + 2 S\'ersics (hosts)): $\chi^{2}_{\nu}$ =  4.167}  \\ \hline
m$_{\rm PSF}$ (AB) & 17.967$\pm$0.001 & 18.882$\pm$0.091 \\
C$_{\rm PSF}$ & 08:23:41.09+24:18:05.5 & 08:23:41.09+24:18:04.9 \\
m$_{\rm host}$ (AB) & 20.400$\pm$0.021 & 18.853$\pm$0.094 \\
$R_{\rm host}$ (kpc) & 5.06$\pm$0.11 & 0.51$\pm$0.07 \\
n$_{\rm host}$ & 0.82$\pm$0.05 & 3.43$\pm$0.40 \\
\hline
\multicolumn{3}{c}{Model 3 (2 PSFs + 2 S\'ersics (hosts) + 1 S\'ersic (lens)): $\chi^{2}_{\nu}$ =  3.636}  \\ \hline
m$_{\rm PSF}$ (AB) & 17.967$\pm$0.001 & 18.916$\pm$0.134 \\
C$_{\rm PSF}$ & 08:23:41.09+24:18:05.5 & 08:23:41.09+24:18:04.9 \\
m$_{\rm host}$ (AB) & 21.632$\pm$0.100 & 19.034$\pm$0.151 \\
$R_{\rm host}$ (kpc) & 5.06$\pm$0.26 & 0.6$\pm$0.1 \\
n$_{\rm host}$ & 0.33$\pm$0.05 & 0.2$\pm$0.28 \\
m$_{\rm lens}$ (AB) &  \multicolumn{2}{c}{ 19.95$\pm$0.021 } \\
C$_{\rm lens}$ &  \multicolumn{2}{c}{ 08:23:41.09+24:18:05.1 } \\
$R_{\rm lens}$ (kpc) &  \multicolumn{2}{c}{3.94$\pm$0.10 } \\
n$_{\rm lens}$  & \multicolumn{2}{c}{1.94$\pm$0.05 }
\enddata
\tablecomments{ 
Best-fit components for 4 different models fit to the F160W image using {\sc galfit}. Central coordinates (C) of model components are in hh:mm:ss.ss+dd:mm:ss.s; the host components are fixed to the same location as the corresponding best-fit PSF. The lensing galaxy component, when present, is denoted with {\it lens} and given as a column between the N and S components. The best-fit magnitude, effective radius, and S\'ersic index for a component are given by m, $R$, and n, respectively.
}
\end{deluxetable}


The residuals of Model 1 suggest that the underlying host galaxies of the quasars are contributing to the extended brightness distributions to some degree, as expected at optical/NIR wavelengths. This is also reflected in the radial profiles where the enclosed brightness is well modeled out to a radius of $\sim$1\arcsec\ (from the center of the N source), but is not well constrained at larger radii. We therefore form Model 2 using the same parameters as in Model 0 with the addition of 2 more S\'ersic profiles to model the underlying N and S host galaxies. We tie the host S\'ersic positions with the corresponding PSFs to reduce model degeneracy (leaving the host locations as free parameters does not improve the fit). We omit the lens galaxy component in Model 2. While this model shows an improvement from the previous two, the residuals suggest that even with the host galaxies included there is some amount of lingering light between the two quasars in the form of a compact clump detected at significance $>10\sigma$. The position of the clump seems consistent with the best-fit location of the lens galaxy component in Model 1. 

To address this clump, we form Model 3 as a combination of Models 1 and 2: Model 3 includes a PSF and S\'ersic profile for both sources plus a central lens galaxy S\'ersic, shown in Figure \ref{fig:HST_galfit}. This model achieves the lowest reduced $\chi^{2}$ statistic out of the four. While the model still leaves some residuals, they are not distributed in a coherent way warranting an additional model component. The residuals also are not detected at more than 50\% above the noise level, so we do not consider them robust indications of merger-induced tidal structures.

\begin{figure*}
     \centering
         \includegraphics[width=\textwidth]{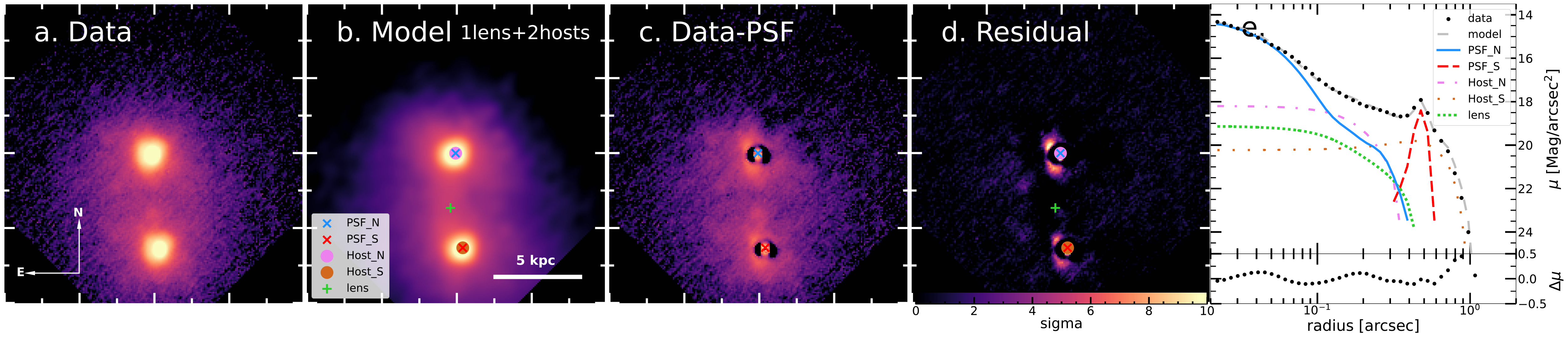}
        \caption{Same as Figure \ref{fig:HST_galfit}, but for Keck AO imaging. We again show Model 3. The extended host components appear more prominent in the $K_{p}$-band, but do not contribute much to the radial profile beyond a radius of 1\arcsec. The irregular shape of the PSF is due to the adaptive optics of Keck. Major tick marks in all panels are 0.25\arcsec. In panel $c$, the faint blob above the S PSF could be a faint foreground galaxy. There also appears to be an arc to the residual structure of the N galaxy that is mirrored in the S galaxy. This overall ellipsoid arrangement suggests some degree of lensing shear.}
     \label{fig:Keck_galfit}
\end{figure*}

The best-fit model (Model 3) characterizes the N quasar as markedly brighter than its S counterpart, with a flux ratio of North to South of $\frac{f_{\rm N}}{f_{\rm S}}\sim2.4\pm0.4$. The fit results paint a different picture for the host galaxies, where the N host galaxy is substantially fainter than the S host (flux ratio of 0.09$\pm0.07$). The N host is much more spatially extended than the S host, which is at the lower limit for the fit parameter such that it functions like a PSF profile. In the radial profiles for both Models 2 or 3, the S host galaxy component appears to mimic the profile of the PSF which results in suppressing the contribution of the PSF, and thus a higher N/S flux ratio. In Model 3 particularly, the S host radius and S\'ersic index approach the minimum allowable values for the fit, which are unphysical. This is likely due to model degeneracy between the PSF and S\'ersic components in the fainter S source, especially when the nearby lens component is added. As a check of whether the S host galaxy is indeed detected, we run the fitting without its inclusion resulting in a substantially poorer fit ($\chi^{2}_{\nu}=5.496$) than most of the other models. We consider the S host galaxy a necessary model component, but limitations of the fitting prevent a detailed comparison of the host galaxy properties. We therefore also consider the results of Model 1 when discussing the flux ratio of the N and S quasars; we obtain a value of $\frac{f_{\rm N}}{f_{\rm S}}\sim1.36\pm0.06$ which is consistent with the previously found flux ratios for photometry using WFC3 filters F475W (1.43$\pm$0.07) and F814W (1.34$\pm$0.07). At the spatial resolution of $HST$, it is difficult to say definitively whether the centrally located brightness distribution is due to a central lens galaxy or the extended underlying galaxies. However, based on the penalized log-likelihood ratio test of the Akaike information criterion \citep[AIC; $e.g.$,][]{Akaike74,Ruffio19}, Model 3 is strongly statistically preferred to both Models 1 and 2, indicating that all of the components are warranted.


\begin{deluxetable}{lcc}
\tabletypesize{\small} \tablewidth{8pt}
\tabletypesize{\scriptsize}
\tablewidth{\textwidth}

\tablecaption{Keck ($K_{p}$) Spatial Modeling
\label{tab:Keck_gal}}
\tablehead{ 
\colhead{Parameter} & \colhead{\obj\ N} & \colhead{\obj\ S}  
}
\startdata 
\multicolumn{3}{c}{Model 0 (2 PSFs): $\chi^{2}_{\nu}$ =  3.581}  \\ \hline
m$_{\rm PSF}$ (AB) & 18.413$\pm$0.002 & 18.644$\pm$0.002 \\
C$_{\rm PSF}$ & 08:23:41.13+24:18:06.0 & 08:23:41.10+24:18:05.4 \\
\hline
\multicolumn{3}{c}{Model 1 (2 PSFs + 1 S\'ersic (lens)): $\chi^{2}_{\nu}$ =  1.295}  \\ \hline
m$_{\rm PSF}$ (AB) & 18.484$\pm$0.001 & 18.749$\pm$0.001 \\
C$_{\rm PSF}$ & 08:23:41.13+24:18:06.0 & 08:23:41.10+24:18:05.4 \\
m$_{\rm lens}$ (AB) &  \multicolumn{2}{c}{ 17.985$\pm$0.004 } \\
C$_{\rm lens}$ &  \multicolumn{2}{c}{ 08:23:41.11+24:18:05.7 } \\
$R_{\rm lens}$ (kpc) &  \multicolumn{2}{c}{4.27$\pm$0.03 } \\
n$_{\rm lens}$  & \multicolumn{2}{c}{0.46$\pm$0.01 }\\
\hline
\multicolumn{3}{c}{Model 2 (2 PSFs + 2 S\'ersics (hosts)): $\chi^{2}_{\nu}$ =  1.211}  \\ \hline
m$_{\rm PSF}$ (AB) & 18.525$\pm$0.001 & 18.808$\pm$0.002 \\
C$_{\rm PSF}$ & 08:23:41.13+24:18:06.0 & 08:23:41.1o+24:18:05.4 \\
m$_{\rm host}$ (AB) & 19.324$\pm$0.008 & 17.842$\pm$0.024 \\
$R_{\rm host}$ (kpc) & 1.91$\pm$0.02 & 5.87$\pm$0.16 \\
n$_{\rm host}$ & 0.51$\pm$0.02 & 1.65$\pm$0.04 \\
\hline
\multicolumn{3}{c}{Model 3 (2 PSFs + 2 S\'ersics (hosts) + 1 S\'ersic (lens)): $\chi^{2}_{\nu}$ =  1.106}  \\ \hline
m$_{\rm PSF}$ (AB) & 18.524$\pm$0.001 & 18.795$\pm$0.001 \\
C$_{\rm PSF}$ & 08:23:41.13+24:18:06.0 & 08:23:41.10+24:18:05.4 \\
m$_{\rm host}$ (AB) & 19.628$\pm$0.011 & 17.925$\pm$0.024 \\
$R_{\rm host}$ (kpc) & 2.25$\pm$0.02 & 6.82$\pm$0.21 \\
n$_{\rm host}$ & 0.29$\pm$0.02 & 1.35$\pm$0.04 \\
m$_{\rm lens}$ (AB) &  \multicolumn{2}{c}{ 19.877$\pm$0.014 } \\
C$_{\rm lens}$ &  \multicolumn{2}{c}{ 08:23:41.11+24:18:05.7 } \\
$R_{\rm lens}$ (kpc) &  \multicolumn{2}{c}{3.19$\pm$0.04 } \\
n$_{\rm lens}$  & \multicolumn{2}{c}{0.20$\pm$0.02 }
\enddata
\tablecomments{ 
Best-fit components for 4 different models fit to the $K_{p}$ image using {\sc galfit}. Central coordinates (C) of model components are in hh:mm:ss.ss+dd:mm:ss.s; the host components are fixed to the same location as the corresponding best-fit PSF. The lensing galaxy component, when present, is denoted with {\it lens} and given as a column between the N and S components. The best-fit magnitude, effective radius, and S\'ersic index for a component are given by m, $R$, and n, respectively.}
\end{deluxetable}


We also model the Keck $K_{p}$-band AO image using {\sc galfit}. For consistency, we adopt the same model components as we used for the F160W image above. The results of the 4 model fits are given in Table \ref{tab:Keck_gal}. We show results of Model 3 in Figure \ref{fig:Keck_galfit}. Here we see stronger indications of underlying disk structure because of the higher S/N compared to the \hst\ image. The disks appear to be aligned in a parallel geometry which suggests a lensing scenario. As with the F160W image, Model 3 (2 PSF point sources, 2 host galaxies, and a lens galaxy) achieves the lowest reduced $\chi^{2}$ value ($\sim$1.1); however, we do not see the same degeneracy between the S source's PSF and S\'ersic components here, as was the case above. We do not detect any faint extended structure in the residual image suggestive of tidal tails. For the quasars, we find a qualitatively similar result to the F160W fit, where the N quasar is brighter than the S quasar. The flux ratio of 1.284$\pm0.002$ for the $K_{p}$-band quasars is roughly consistent with the values above. Here, we are quoting the flux ratio derived from Model 3, although the value derived from Model 1 (1.276$\pm$0.002) is also roughly consistent. The flux ratio of the host galaxies is less extreme (0.208$\pm0.025$) than found for the $HST$ image. However, the best-fit sizes of host components appear to be reversed from the F160W fitting above, such that the S host here is much more extended (beyond the physical separation of the two nuclei) compared to the N host; this once again suggests the limitations of our spatial analysis with such a blended pair and model degeneracy.

Because of the lower sensitivity of Keck compared to $HST$, it is not as obvious in panel $c$ of Figure \ref{fig:Keck_galfit} whether there is in fact a cluster of bright residuals constituting a central lens galaxy; however, there appears to be a faint blob detected at $\sim4\sigma$ significance towards the S source that is spatially consistent with the location of the lens galaxy in the $HST$ image modeling, and its inclusion is statistically warranted. As with the \hst\ modeling, the AIC values indicate that Model 3 is strongly statistically preferred over Models 1 and 2, bolstering the need for including both the lensing galaxy and the host galaxy components.

\subsection{Optical/IR Spectral Fitting}\label{sec:qsofit}
\begin{figure*}
     \centering

     \begin{subfigure}
         \centering
         \includegraphics[width=\textwidth]{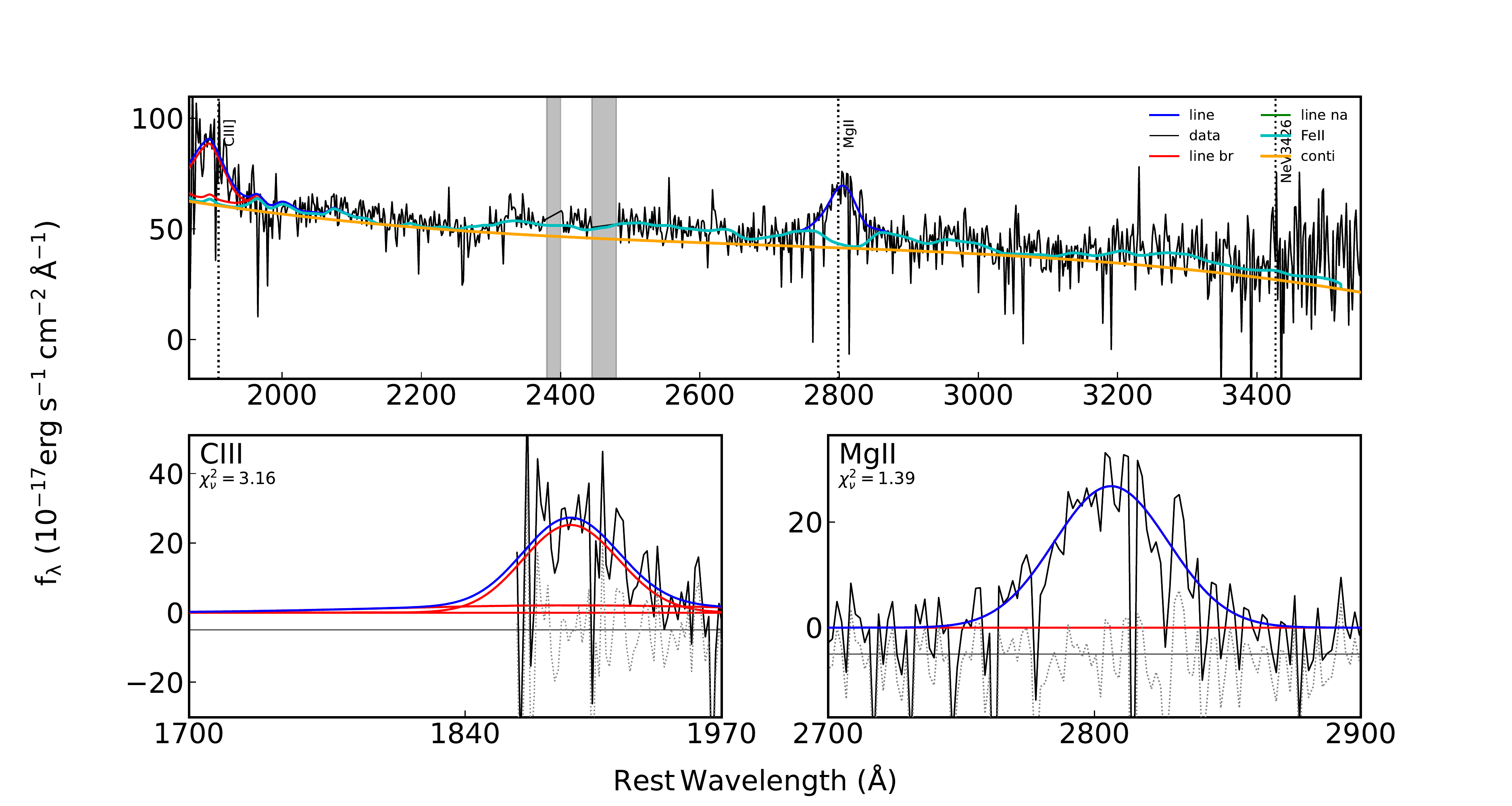}
     \end{subfigure}
     \hfill
     \begin{subfigure}
         \centering
         \includegraphics[width=\textwidth]{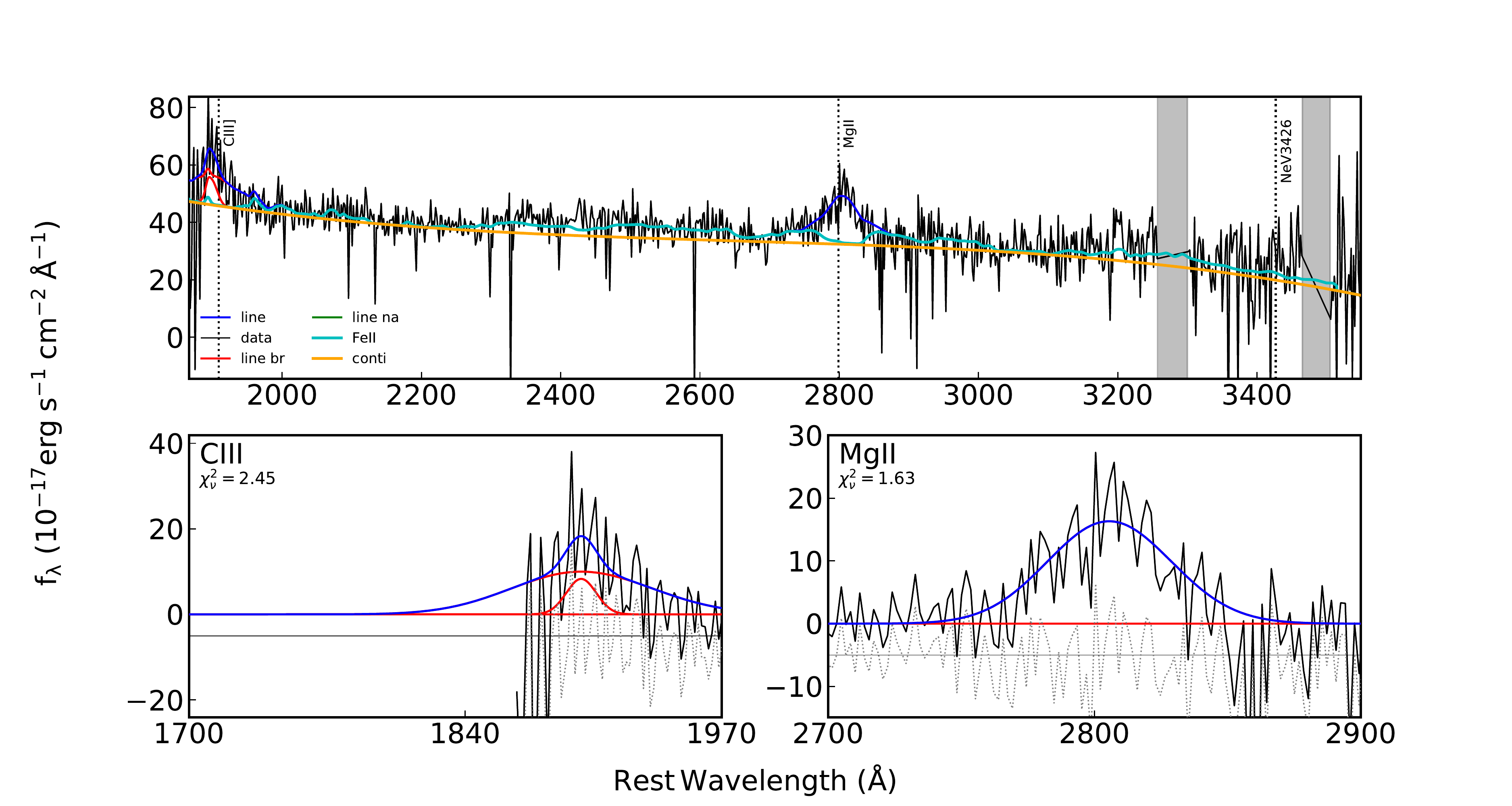}
     \end{subfigure}
        \caption{Spectral fitting of the $HST$/STIS spectra for \obj. The top and bottom sets of panels show the fits for the N and S targets, respectively. The spectra are fit with a pseudo-continuum model composed of a power law combined with a low-order polynomial (orange curve) and the \FeII\ template. Gray regions are masked out during the fit due to lingering cosmic rays. The lower inset panels show zoom-ins on the local fits of broad emission lines after continuum subtraction.  }
     \label{fig:STIS}
\end{figure*}

At first glance, the N and S components appear to be similar, with a nearly constant flux ratio. The flux calibration of the GMOS spectra yields systematically lower fluxes compared to the STIS spectra. Since both nuclei appear equally affected, it does not influence our bulk results of comparing the spectra for differences. To check for differences, we normalize the GMOS spectrum of the S source by the average flux ratio observed between the two sources (shown with dashed line in Figure \ref{fig:opspec}), and then take the difference to reveal any obvious unique features. The GMOS spectra are remarkably similar, apart from several isolated features, some of which are artefacts. On the other hand, the STIS spectra show several differences between the N and S sources. Subtle variations in the broad shapes of the STIS spectra might be a result of the poor S/N.  In particular, the STIS-N spectrum appears to exhibit two strong narrow emission lines around $\lambda_{\rm obs} = 6895 {\rm \ and\ } 6715$ \AA\AA. Given the redshift of the object, these two features do not correspond to any typical strong quasar emission lines. Considering that we see a similar feature in the STIS-S spectrum at $\sim9200$\AA, we believe these features are due to uncleaned cosmic rays, and so we have masked them in Figure \ref{fig:opspec}, as well as with gray regions in Figure \ref{fig:STIS}. The STIS-S spectrum appears to have absorption features not present in the N spectrum. These disparate features could be indications that the two spectra originate from two non-lensed quasar cores. However, we do not see evidence of these same features in the corresponding GMOS spectra, which is much less noisy than the STIS observations. An important caveat here is that the GMOS extraction regions are not as well separated as those used for STIS as a consequence of the PSF and seeing, so it is difficult to conclude whether those features are truly absent in the GMOS spectra or have been washed out due to some degree of blending between both sources in each individual spectrum. 

\begin{figure*}
     \centering

     \begin{subfigure}
         \centering
         \includegraphics[width=\textwidth]{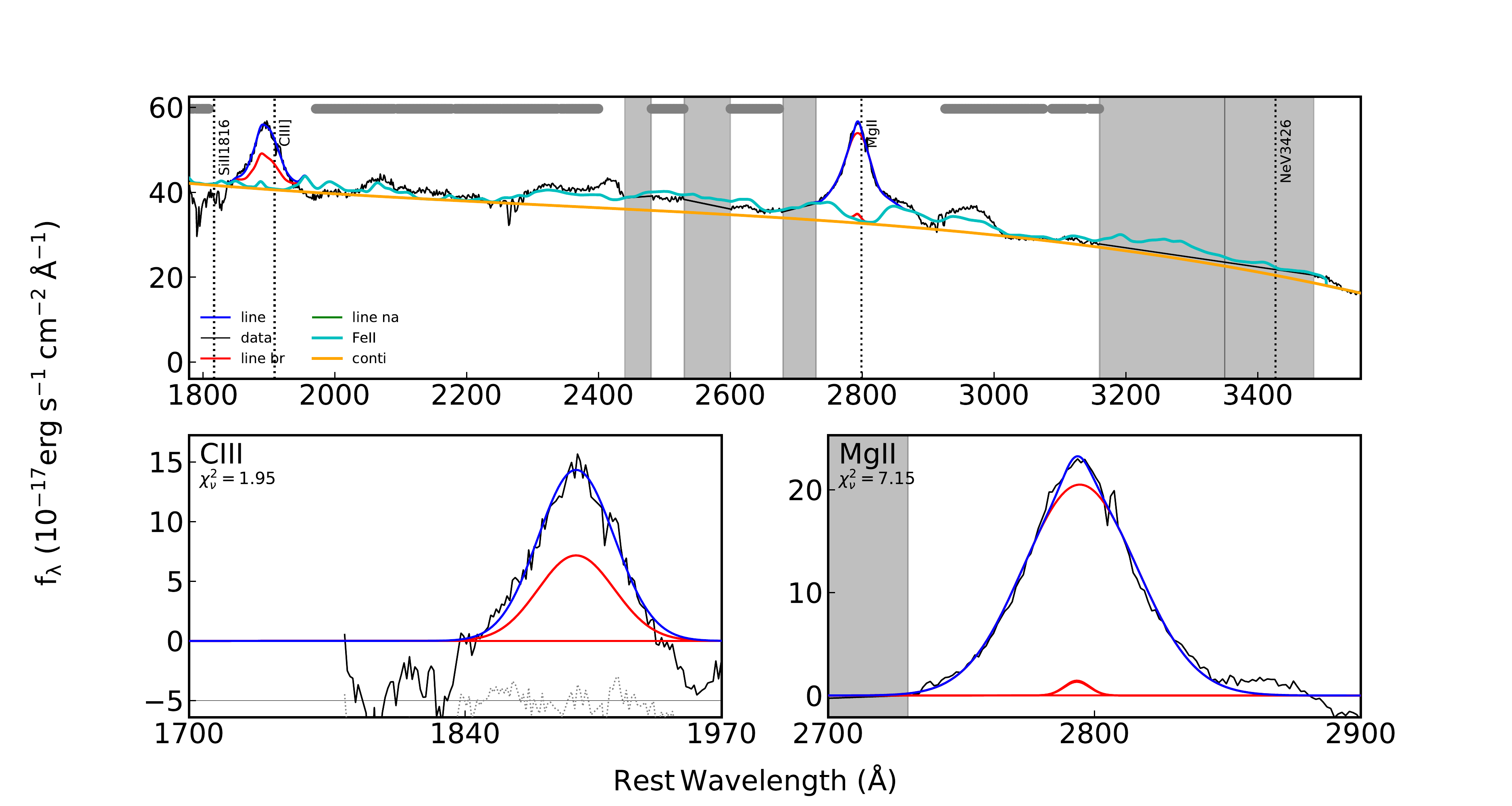}
     \end{subfigure}
     \hfill
     \begin{subfigure}
         \centering
         \includegraphics[width=\textwidth]{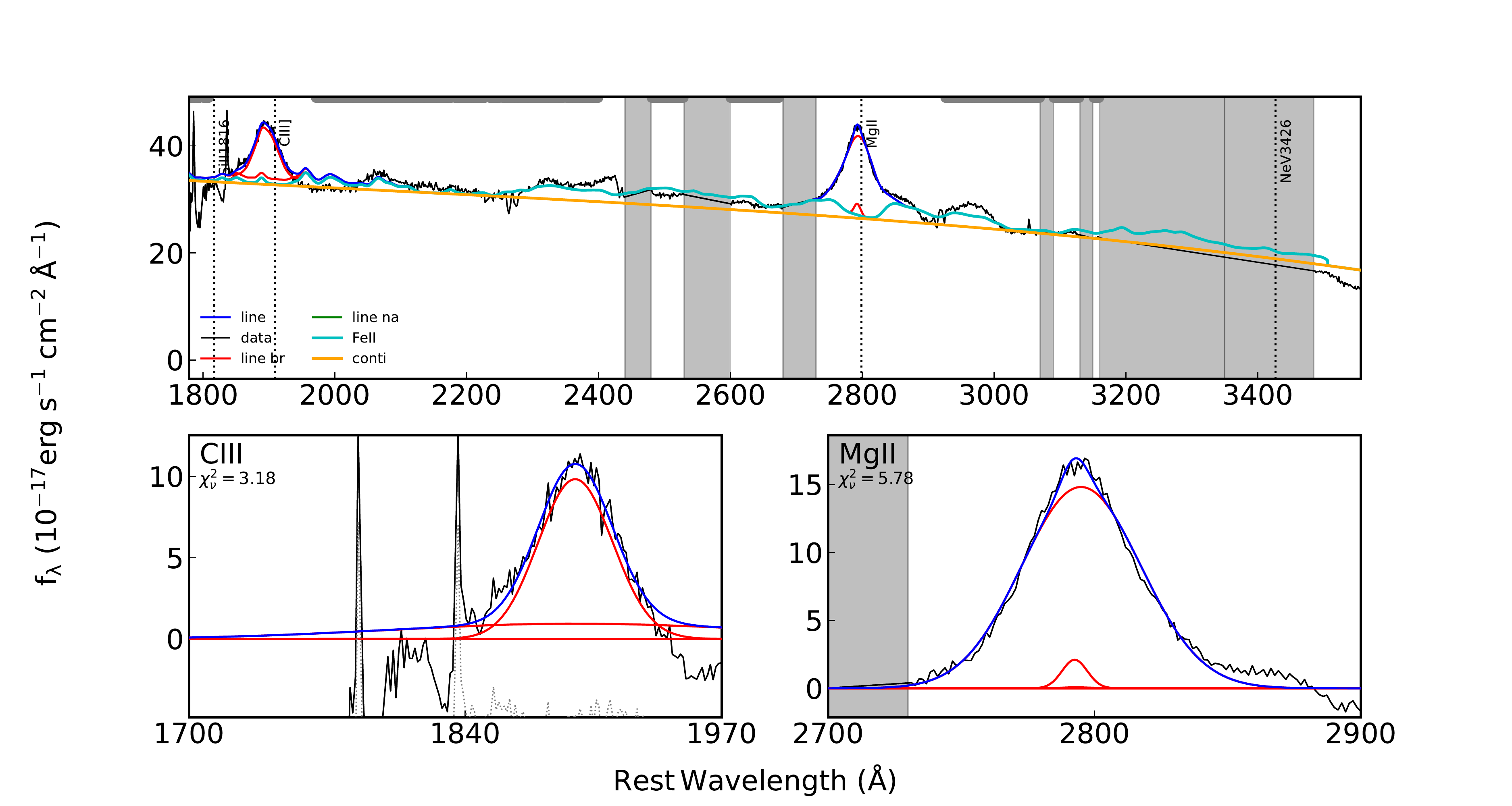}
     \end{subfigure}
        \caption{Same as Figure \ref{fig:STIS}, but spectral fitting of the Gemini/GMOS spectra. Gray regions are masked out primarily to avoid regions that are heavily affected by telluric absorption bands. The extended range towards bluer wavelengths allows for a better fit of the C{\sc iii]} emission line than is possible with the STIS spectra.}
     \label{fig:GMOS}
\end{figure*}

To constrain the properties of the quasars' emission lines and the underlying continua, we employ the spectral fitting code PyQSOFit \citep{Guo18}. We model each spectrum separately because of differences in spectral resolution. We mask out regions with persistent artefacts of the spectral reduction. The spectra are first corrected for extinction using the Milky Way extinction law of \citet{Cardelli89} and drawing upon the dust maps from \citet{Schlegel98}, and then de-redshifted to rest-frame. We model each galaxy as a pseudo-continuum that is a linear combination of a power law, \FeII\ emission, and a low-order polynomial for quasar continuum, combined with Gaussian components for broad and narrow emission lines. Local fits are then done for the emission line regions after subtracting the continuum elements. We show the spectral fits for the STIS and GMOS data in Figures \ref{fig:STIS} and \ref{fig:GMOS}, respectively. 

While we obtain reasonable fits for the GMOS spectra, there are several windows where residual telluric absorption features need to masked out; one of these windows truncates the broad \MgII\ emission line, and so we rely predominantly on the fit results of the STIS spectra for this emission line fit. Uncertainties in fitted parameters are calculated using Markov Chain Monte Carlo simulations. We determine the systematic redshifts of the nuclei based on the velocity offsets observed for the emission lines. The best-fit redshifts are 1.81354$\pm0.00036$ and 1.81328$\pm0.00036$ for the N and S nuclei, respectively, which agree within the errors. We determine the slopes of the underlying continua to be -1.030$\pm$0.014 and -1.031$\pm$0.005 for the N and S sources. The slopes are consistent, suggesting that the quasars are the same object doubly imaged. 

\begin{deluxetable}{lcr}
\tabletypesize{\small} \tablewidth{9pt}
\tablewidth{\textwidth}

\tablecaption{Emission Line Properties of \obj\
\label{tab:emi_lines}}
\tablehead{ 
\colhead{Emission Line} & \colhead{\obj\ N} & \colhead{\obj\ S}  \\
\colhead{Measurement} & \colhead{} & \colhead{}  
}
\startdata 
C{\sc iii}{]} Flux & 692.1 $\pm$ 8.2   & 688.1 $\pm$ 33.1  \\\vspace{0.15cm}
\MgII\ Flux & 1433.8 $\pm$ 59.7  & 949.8 $\pm$ 58.4  \\
C{\sc iii}{]} EW & 17.0 $\pm$ 0.2  & 21.0 $\pm$ 1.0  \\\vspace{0.15cm}
\MgII\ EW & 34.6 $\pm$ 1.4  & 29.3 $\pm$ 1.8  \\
C{\sc iii}{]} FWHM & 7142.1 $\pm$ 99.5 & 7424.8 $\pm$ 162.6 \\
\MgII\ FWHM & 5400.4 $\pm$ 268.4 & 5871.2 $\pm$ 560.9  \\
\enddata
\tablecomments{ 
Emission line properties measured from the spectral fitting of the $HST$/STIS and Gemini/GMOS spectra in the observed-frame optical. Flux is reported in units of 10$^{-17}$ erg s$^{-1}$ cm$^{-2}$, rest-frame equivalent width (EW) is in units of \AA, full width at half max (FWHM) is in units of km s$^{-1}$. All values are for the broad component of the line. All errors are quoted at the 1-$\sigma$ level from Monte Carlo simulations.
}
\end{deluxetable}
The resulting best-fit parameters of the UV emission lines are given in Table \ref{tab:emi_lines}. The fluxes of the \MgII\ lines are different between the two sources by $\sim$50\%, which is again a roughly consistent flux ratio to other wavebands. The equivalent width and FWHM values are roughly consistent. Because the C{\sc iii}] emission line is not fully covered by the STIS spectra, we rely on the GMOS spectral fitting for measurements in this case. Again, we find that the line properties are roughly consistent between the N and S nuclei. The flux ratio of the continuum at rest-frame 2500 \AA\ is $\sim$1.35 which is consistent with the results above from photometry.

We estimate the black hole mass for each nucleus using the empirical mass estimator from \citet{Vestergaard09} using \MgII\ emission:
\begin{equation}
M_{\rm BH} = 10^{6.86}\left[\frac{{\rm FWHM(MgII)}}{{\rm 1000\ km\ s^{-1}}} \right]^{2}\left[ \frac{{\rm \lambda} L_{\rm \lambda}(3000\ {\rm \AA})}{10^{44}\ {\rm erg\ s^{-1}}}\right]^{0.5}\ M_{\odot} ,
\end{equation}
where ${\rm \lambda} L_{\rm \lambda}$ is the monochromatic luminosity taken at 3000 \AA\ obtained from our continuum fitting. We list the resulting black hole masses in Table \ref{tab:phot}. Owing to the similarity of the spectra, the derived masses are consistent with each other. The $\sim$50\% differences in \MgII\ flux, which could be due to lensing magnification, do not impact the derived masses as much as the FWHM values, which are consistent to within 9\%.    

\begin{deluxetable*}{lcccccc}
\tabletypesize{\small} \tablewidth{9pt}
\tablewidth{\textwidth}

\tablecaption{Radio Properties of \obj\
\label{tab:VLA}}
\tablehead{ 
\colhead{Radio Designation} & \colhead{$S^{\rm peak}_{\nu}$} & \colhead{$S^{\rm int}_{\nu}$}  & \colhead{log $\nu L_{\nu}$} & \colhead{Maj} & \colhead{Min} & \colhead{PA} \\
\colhead{J2000}  & \colhead{($\mu$Jy/beam)} & \colhead{($\mu$Jy)}& \colhead{(erg s$^{-1}$)}  & \colhead{(arcsec)} & \colhead{(arcsec)} & \colhead{(deg)} \\
\colhead{(1)} & \colhead{(2)} & \colhead{(3)} & \colhead{(4)} & \colhead{(5)} & \colhead{(6)} & \colhead{(7)}  
}
\startdata 
\multicolumn{7}{c}{C-Band ($\nu$ = 6 GHz)} \\
082341.08$+$0241805.7 & 335.8$\pm$6.4 & 365.0$\pm$14.0 & 42.131$\pm$0.001 & 0.333$\pm$0.105 & 0.069$\pm$0.034 & 106$\pm$11\\
082341.08$+$0241805.0 & 237.1$\pm$6.4 & 254.0$\pm$13.0 & 41.556$\pm$0.002 &  $<$0.969$\pm$0.043 & $<$0.357$\pm$0.006 & 112$\pm$1 \\ 
\multicolumn{7}{c}{Ku-Band ($\nu$ = 15 GHz)} \rule{0pt}{4ex} \\
082341.08$+$0241805.7 & 61.5$\pm$6.9 & 60.0$\pm$12.0 & 41.745$\pm$0.009 & $<$0.137$\pm$0.019 & $<$0.080$\pm$0.007 & 160$\pm$7\\
082341.08$+$0241805.0 & 59.6$\pm$7.9 & 98.0$\pm$19.0 & 41.54$\pm$0.005 &  0.113$\pm$0.027 & 0.046$\pm$0.033 & 103$\pm$21
\enddata
\tablecomments{ 
(1) J2000 coordinate of the identified radio source core, where we list the N nucleus first;
(2) Peak flux density at the central frequency of the band within the source extent;
(3) Integrated flux density at the central frequency of the band within the source extent;
(4) logarithm of the rest-frame luminosity density at the central frequency of the band; 
(5$-$8) Best fit beam-deconvolved source sizes (FWHMs in arcsec) along the major and minor axes and the position angle of the major axis (degrees east of north). In cases of unresolved point sources (denoted with $<$), the values for are for the un-deconvolved regions. All errors quoted represent 1-$\sigma$ uncertainties which are derived via the correlated noise prescription of \citet{Condon97}. The error of the photometry does not include the 3\% uncertainty in the VLA flux density scale \citep{Perley13}. 
}
\end{deluxetable*}

\subsection{Radio Imaging}\label{sec:radio_img}

\begin{figure*}
     \centering
         \includegraphics[width=\textwidth]{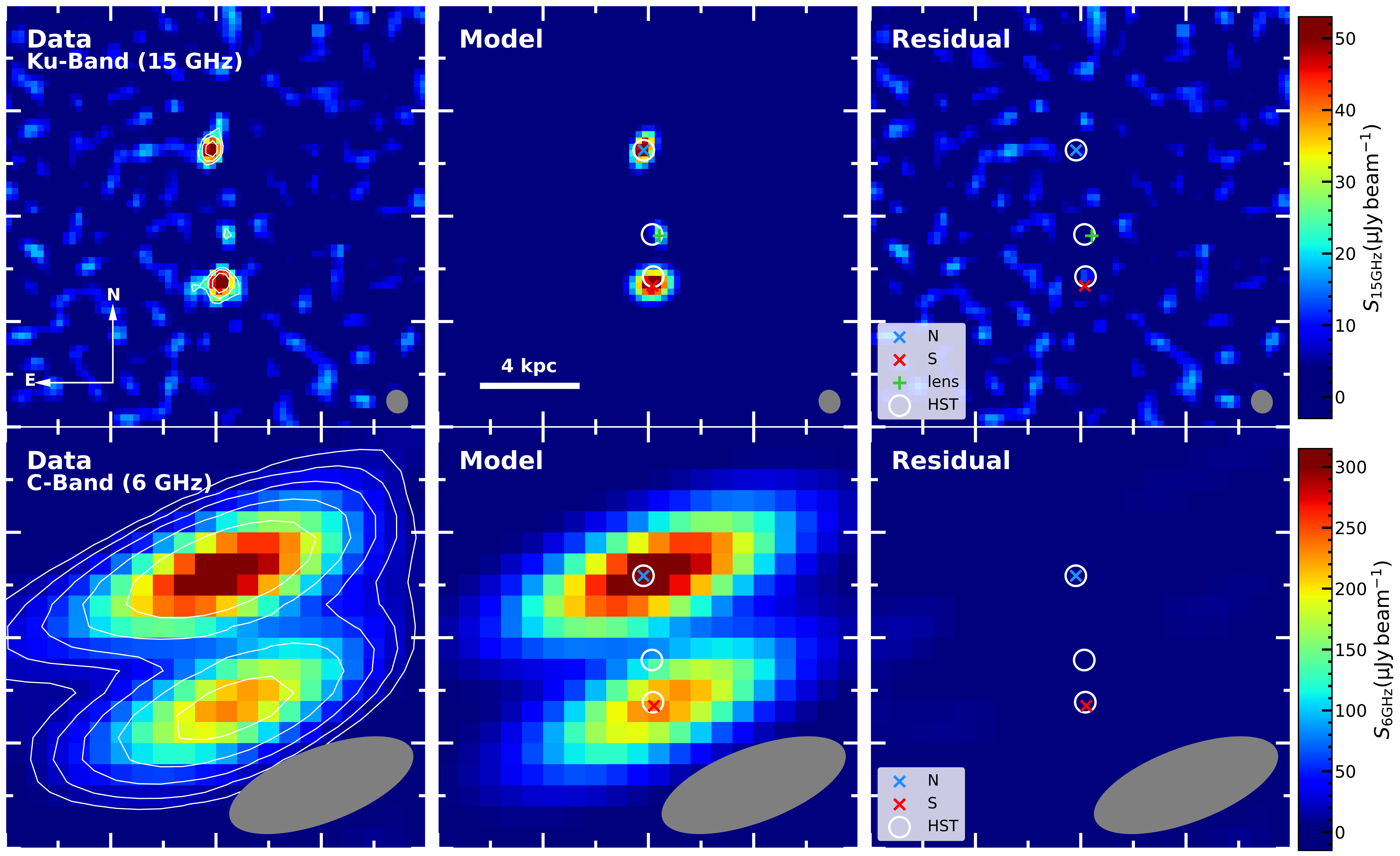}
        \caption{{\bf Top row:} VLA Ku-band image decomposition of \obj\ at 15 GHz. We show from left to right: the continuum image, the model using \texttt{imfit}, and the residual map. The restoring beam is shown in gray in the bottom right corner of each panel. Contours begin at 3 times the rms background and increase exponentially to the peak value. The legend in the right panel indicates the N and S radio components, as well as the compact central blob that might correspond to the foreground lensing galaxy. We overlay the $HST$ F160W best-fit coordinates, aligned by the N source, which show agreement to within 0.1\arcsec. Major tick marks are spaced by 0.5\arcsec. {\bf Bottom row:} same as the top row, but for the VLA C-band continuum imaging at 6 GHz. The best-fit locations between the two bands are consistent to within 0.1\arcsec. While the N source is clearly brighter and more spatially extended at 6 GHz, the reverse situation is seen in the 15 GHz image, indicating different spectral slopes between the N and S quasars. We do not find evidence for a lensing galaxy component in the 6 GHz frame}
     \label{fig:VLA}
\end{figure*}
We use the CASA task \texttt{imfit} to quantify the spatial extents and flux densities of the sources, simultaneously fitting with a Gaussian component for each source. The resulting best-fit quantities are given in Table \ref{tab:VLA}. Figure \ref{fig:VLA} shows the cleaned imaging, model, and residual maps for both the C and Ku-band data. The spatial extents are slightly larger than the restoring beam sizes, and show an interesting dichotomy: the N source appears more spatially extended at 6 GHz while the S source is more extended at 15 GHz. This could be suggestive of two separate quasars. Specifically, we find that \obj\ N is unresolved in the Ku-band image (beam FWHM: 0.11$\times$0.10\arcsec), while \obj\ S is only marginally resolved in the C-band image (beam FWHM: 0.92$\times$0.35\arcsec). This is reflected in the markedly different integrated flux values, and thus the spectral indices, between the two sources. \obj\ N has $\alpha_{\rm 6-15\ GHz}$ = $-$1.971$\pm$0.222, while \obj\ S has an index of $\alpha_{\rm 6-15\ GHz}$ = $-$1.039$\pm$0.219. Both sources exhibit steep negative values consistent with non-thermal optically-thin synchrotron emission. The observed-frame flux ratio between the N and S source is 1.437$\pm$0.064 at 6 GHz, which is consistent with that measured in the optical and NIR. However, due to the disparity of the spectral indices, the flux ratio at 15 GHz is 0.612$\pm$0.280, which seems to argue against a lensing scenario.

The Ku-band image also shows an intriguing central clump of residuals. At first glance, its location appears to be roughly consistent with the putative foreground lensing galaxy as seen in the Keck imaging. We add an elliptical component during the model fitting to incorporate this blob, as shown in the top middle panel of Figure \ref{fig:VLA}. The additional component does not alter the best-fit solutions for the N and S sources, and can be successfully modeled. However, this unresolved central source is slightly smaller than the restoring beam which suggests that it could be noise, and it has an integrated flux density consistent with the background within its uncertainties yielding a detection at only the 2-$\sigma$ level. 

The faintness of the blob is not surprising giving the faintness of the lens component in the NIR bands; the observed low level of radio emission is likely not due to an AGN, although could be attributed to star formation processes. A component between the N and S sources cannot be successfully modeled for the C-band image; adding the additional component results spurious best-fit positions and non-physical flux values. While it is possible that a faint source does exist between the two sources, they are blended at the location where the additional fainter source would be located.

\subsection{X-ray Spectral Fitting}\label{sec:BAYMAX}
Given the spatial resolution of $Chandra$ ($\sim$0.5\arcsec native pixel size) and the fact that the \obj\ system appears somewhat blended even in the higher-resolution IR imaging, we begin our analysis by confirming that there are indeed two X-ray sources detected. To disentangle the sources from the detector effects, we utilize the Bayesian AnalYsis of Multiple AGN in X-rays ({\sc {\sc baymax}}) algorithm, which is described in detail in \citet{Foord19, Foord20}. Briefly, Baymax returns the likelihood of the two-source model over the one-source model via the Bayes factor $\mathcal{B}$, and probabilistically assigns each X-ray event to one of the sources in the two-source model.

\begin{figure*}
     \centering
    \includegraphics[width=\textwidth]{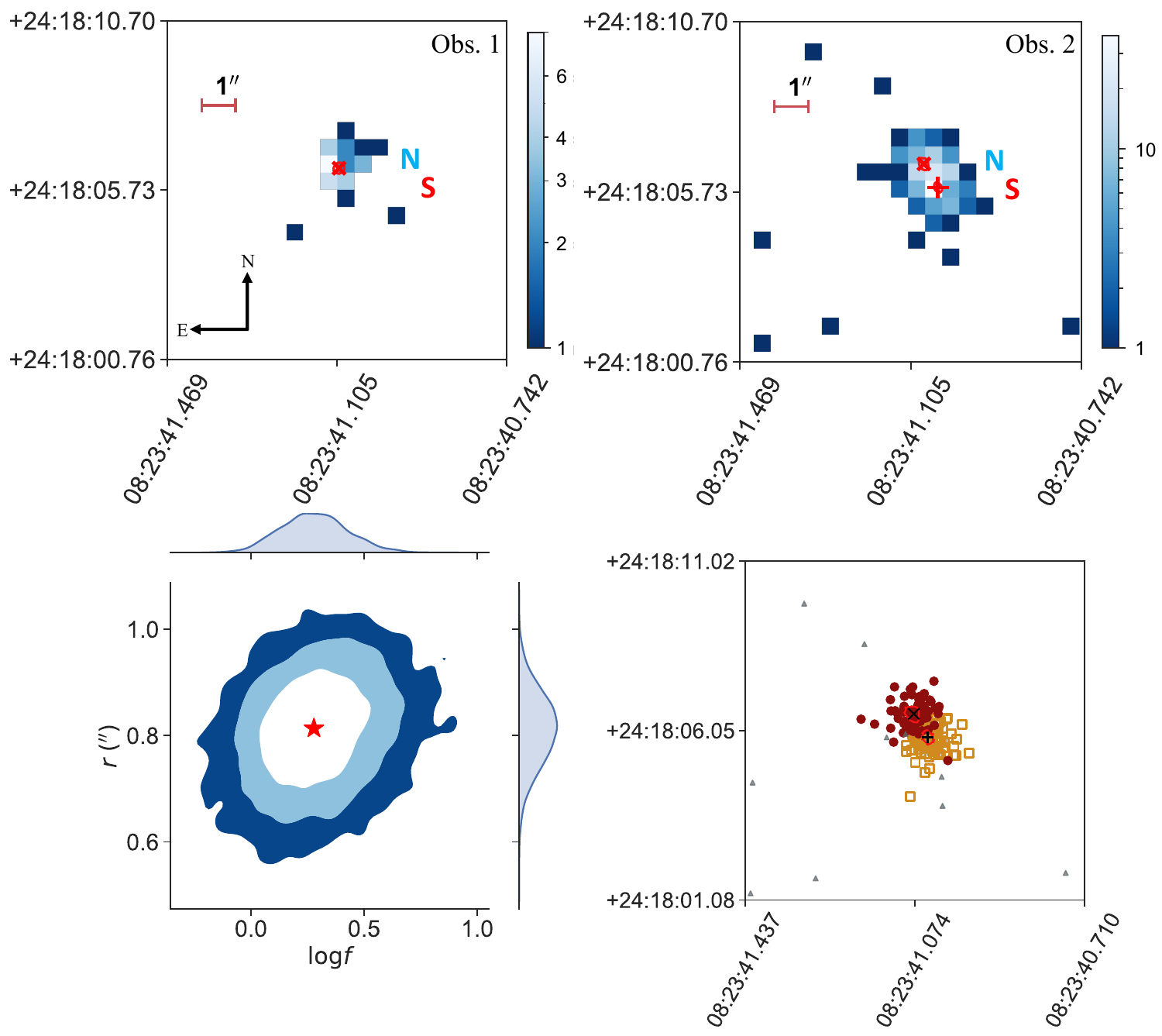}

        \caption{Chandra X-ray observations and image decomposition. {\bf Top row:} Imaging of observations 1 and 2 (left and right panels) for \obj\ where the color bar indicates observed broad-band counts. In observation 1, no source is detected for the S quasar. In observation 2, we mark the best-fit locations of the two sources from {\sc baymax}. The location of the brightest pixels has shifted noticeably between the two observations. {\bf Bottom left:} {\sc baymax} assessment of joint posterior distribution for separation and flux ratio of two sources which is roughly consistent with our {\sc galfit} modeling results. {\bf Bottom right:} Decomposition of the broad-band counts for observation 2, assigning each count to the most likely source.}
     \label{fig:chandra}
\end{figure*}

In Figure \ref{fig:chandra}, the counts are color-coded based on whether they are most likely originating from either the N or S source. For both observations, we restrict {\sc baymax} to a 20\arcsec$\times$20\arcsec\ region centered on \obj. While visually obvious, we confirm that observation 1 is more consistent with a single source model with $ln(\mathcal{B}) = 0.6\pm1.4$, where all 33 counts are associated with the N source. For observation 2, we obtain $ln(\mathcal{B}) = 17.3\pm1.8$, which is strongly in favor of two source model. Here, 57 counts are associated with the N source, and 103 are associated with the S source. The uncertainty on $\mathcal{B}$ is the 68\% confidence interval, taken from the statistical errors on $\mathcal{B}$, returned from the \texttt{nestle} package. 

In Figure \ref{fig:chandra}, we show the posterior distribution of the separation ($r$) and count ratio ($f$) derived by {\sc baymax} via MCMC sampling for observation 2. We find a median separation of 0.81\arcsec$^{+0.19}_{-0.22}$, which is consistent within the quoted 3$\sigma$ uncertainties with the separation determined from Gaia. However, the median count ratio between the weaker N source and the  dominant S source during observation 2 is 0.51$\pm0.56$, which is quite different from the flux ratios determined above. 
\begin{figure*}
     \centering
    \includegraphics[width=\textwidth]{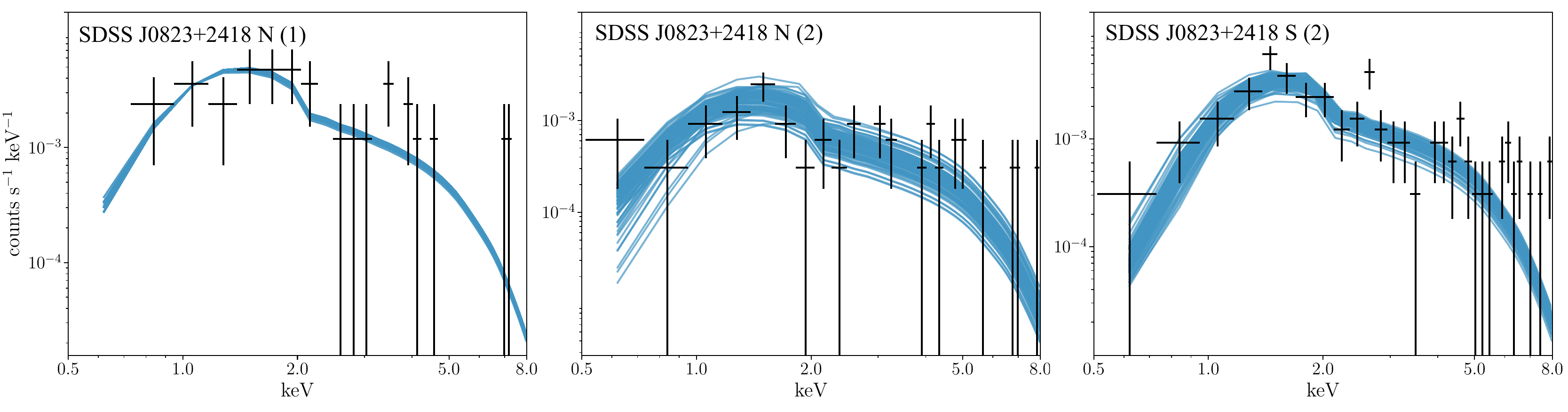}
        \caption{Chandra X-ray spectral fitting for \obj\ using {\sc baymax}. The target is denoted as either North (N) or South (S) in the label, and observation number is given in parentheses. The SW nucleus is not detected in the first observation, and then dominates in the later observation. The spectral shapes are fairly similar, although the N source appears to have a less steep cutoff in the soft X-ray regime than the S source for both observations.}
     \label{fig:chandra2}
\end{figure*}

The extracted spectrum for each source is composed of the counts which have been spatially decomposed by {\sc baymax}, and the spectral fits are evaluated using Cash statistics. The background spectrum is similarly determined for the field (gray points in Figure \ref{fig:chandra}). The spectra are fit using the combined XSPEC model \texttt{phabs * zphabs * zpow}, representing a redshifted powerlaw with photoelectric absorption by local material. Milky Way obscuration for \texttt{phabs} is set to the value of Milky Way neutral Hydrogen column density in the direction \obj\ ($N_{\rm H} = 3.77\times10^{20}$ cm$^{-2}$), found using the {\sc propcolden} tool \citep{Dickey90}. We fix the photon index to the canonical value for AGN of $\Gamma = 1.8$. We did attempt leaving $\Gamma$ as a free parameter, but this consistently resulted in poorer fits. Sampling from the posterior distributions, {\sc baymax} creates 100 spectral realizations for both the N and S source, and then fits them via XSPEC. 

\begin{deluxetable}{lcccc}
\tabletypesize{\small} \tablewidth{9pt}
\tabletypesize{\scriptsize}
\tablewidth{\textwidth}
\tablecaption{X-ray properties of \obj\
\label{tab:x-ray}}
\tablehead{
\colhead{Target} & \colhead{Counts}  &\colhead{$F_{\rm 2-7\ keV}$} &\colhead{HR}   &\colhead{$L_{\rm 2-7\ keV}$} \\
\colhead{(1)}& \colhead{(2)}& \colhead{(3)}& \colhead{(4)}& \colhead{(5)}
}
\startdata \vspace{0.25cm}
\obj\ N(1) & 33  & 12.0$^{+0.2}_{-0.5}$ & 0.21$^{+0.03}_{-0.12}$   & 13.0$^{+0.6}_{-1.0}$ \\ 
\obj\ N(2) & 57  & 4.5$^{+2.4}_{-2.1}$ & 0.38$^{+0.12}_{-0.24}$   &  4.6$^{+4.1}_{-2.1}$ \\
\obj\ S(2) & 102.5 & 8.6$^{+2.0}_{-2.6}$ & 0.54$^{+0.05}_{-0.1}$  &  12.0$^{+3.0}_{-3.0}$ 
\enddata
\tablecomments{ 
X-ray properties derived using \texttt{{\sc baymax}} on the individual $Chandra$ ACIS-S observations. (1) Target and frame of observation (the S source is not detected in observation 1); (2) broad-band net photon counts; (3) observed flux in units of 10$^{-14}$ erg s$^{-1}$ cm$^{-2}$; (4) hardness ratio: HR = (H-S)/(H+S), where H and S are the net counts in the hard and soft X-ray bands, respectively; (5) rest-frame unabsorbed luminosity from best-fit spectral model (assuming fixed power law photon index $\Gamma = 1.8$), in units of 10$^{44}$ erg s$^{-1}$. Values quoted are the medians resulting from 100 spectral realizations, and all errors are quoted at the $\sim$3-$\sigma$ confidence level.     
}
\end{deluxetable}

We show all the spectral realizations for each source and observation in Figure \ref{fig:chandra2}. We give the resulting derived quantities in Table \ref{tab:x-ray}. The average intrinsic luminosities of $L_{\rm X} \sim10^{45}$ erg s$^{-1}$ are securely within the quasar regime. The intrinsic unabsorbed flux ratio between the two sources during observation 2 is $0.38^{+0.35}_{-0.20}$, which is starkly different from the flux ratios found in the optical and UV regimes. The spectra for both observations of the N source exhibit some degree of soft X-ray counts while the S source shows a steeper decline below 2 keV. This is reflected in the median values of the hardness ratio, HR $= \frac{H-S}{H+S}$, where and $S$ and $H$ are the counts in the soft (0.5$-$2 keV) and hard (2$-$7 keV) bands, respectively. The value of HR is consistent within the uncertainties between the two observations for the N source (0.21$^{+0.03}_{-0.12}$ vs. 0.38$^{+0.12}_{-0.24}$), despite its flux diminishing by $\sim$50\%. We find a higher HR (0.54$^{+0.05}_{-0.1}$) for the S source, albeit within the scatter of the N source for observation 2. Were this a bona fide dual quasar, the HR values could be interpreted as a higher degree of obscuration in the S source. Differential extinction could also be due to different lines of sight through the foreground galaxy. For a quadruply lensed quasar, \citet{Glikman23} found low degrees of differential extinction ($<$0.15 mag in NIR/optical), which would also contribute to differential X-ray absorption. Given that the two sources are unobscurred broad line quasars in optical/UV and the previous evidence in support of the lensing scenario, this temporal evolution of the two apparent sources instead hints that perhaps some intrinsic variability of is at play, or a lensing-based time lag, or a combination of both effects.  

\section{Discussion}\label{sec:discuss}
We first combine the findings from the various lines of evidence in our multi-wavelength analysis. Our primary goal is to assess whether \obj\ is  a dual AGN  or if it is a single AGN that has been gravitationally lensed by a foreground galaxy. Our IR image decomposition has shown that inclusion of a central galaxy component between the two sources, even when accounting for the host galaxies, yields the lowest-residual model fits that are likewise statistically preferred. In the Keck image, the location of this central galaxy appears marginally discernible above the background; and while not directly resolved in the HST image, the residual flux centrally concentrated between the two quasars and hosts is significant at $>5\sigma$. There is also tentative evidence of the same central galaxy in the Ku-band radio imaging, although it is not resolvable in the other band. The strikingly similar UV/optical spectra of the two quasars are naturally explained as lensed images of the same source, such that the emission line and continuum properties are in good agreement within their uncertainties. Finally, the optical and IR photometry, as well as the continuum UV luminosities yield consistent flux ratios between the N and S sources indicating that their underlying spectral energy distributions (SEDs) are the same modulo a scaling factor that can be attributed to magnification due to lensing.  

\begin{figure}
     \centering
         \includegraphics[width=0.48\textwidth]{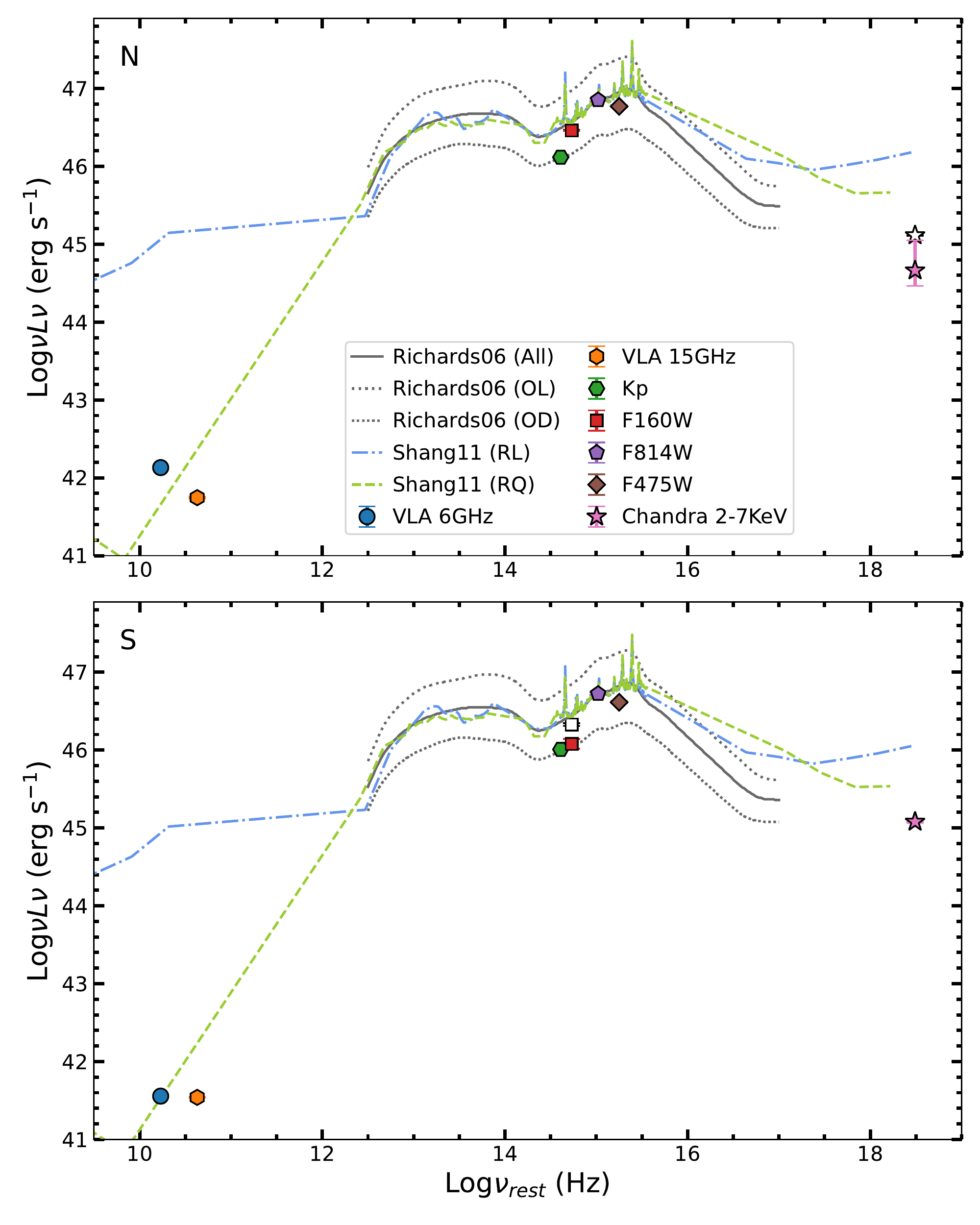}
        \caption{SEDs for SDSS J0823+2418. The top and bottom panels show the SEDs for the N and S source, respectively. We also plot optically luminous (OL), optically dim (OD), and the mean (All) SEDs of optically selected quasars from SDSS as a comparison \citep{Richards06b}. Typical SEDs of optically bright non-blazar radio-loud (RL) and radio-quiet (RQ) quasars are also plotted for reference \citep{Shang11}. The literature SEDs are normalized to the $HST$ F814W luminosities of the target. For the N source, we plot the X-ray luminosity from observation 1 using a colorless star. The pink stars refer to observation 2 luminosities in both panels. For the F160W fluxes, the red squares refer to Model 3, while the colorless squares refer to Model 1 (N value is nearly identical in both models).  }
     \label{fig:sed}
\end{figure}

We plot the SEDs of SDSS J0823+2418 based on our multi-wavelength photometry in Figure \ref{fig:sed}. We overplot several template SEDs of typical quasars from the literature for comparison \citep{Richards06b, Shang11}. The SEDs of the N and S source are qualitatively similar across the rest-frame optical/UV regime, and both show a steeper slope than the template SEDs towards the Keck $K_{\rm p}$ filter. We show the differing values for the S source between models 1 and 3 in the F160W filter, where the Model 1 value produces a flux ratio that is consistent with the other UV/optical bands. Both sources appear to be underluminous in X-ray and radio bands compared to the optically bright quasar templates. The large uncertainty for the X-ray luminosity of the N source during observation 2 renders the value roughly consistent with the value from observation 1.

Since both sources have radio luminosities near the radio-quiet quasar template, we compute the radio loudness parameter, $R_{\rm 6\ cm/ 2500\ \mathring{A}}$, as the ratio of the flux densities at rest-frame 6 cm and 2500 \AA. We scale the observed 6 GHz flux densities to 6 cm, assuming a power law with the spectral index values $\alpha_{\rm 6-15\ GHz}$ observed for each source. Radio-loud quasars are typically defined as having $R_{\rm 6\ cm/ 2500\ \mathring{A}} > 10$ \citep{Kellermann89}. We find that the N source is above this threshold with $R = 19.32 \pm 5.27$, while the S source is radio-quiet with $R = 5.33 \pm 1.57$. This difference is mostly due to the two-band spectral indices. If we instead compute $R$ assuming the canonical value of $\alpha\sim-0.7$ for AGN due to synchrotron emission, then we find that values for the N and S sources ($R = 4.12 \pm 1.12$ and $R = 3.53 \pm 0.97$, respectively) are consistent and definitively in the radio-quiet regime.

Despite the large amount of evidence pointing towards a lensed source, there remain several pieces of evidence which appear to be more suggestive of two distinctive quasars. While the C-band flux ratio is consistent with those at other wavelengths, the higher-resolution Ku-band flux ratio is reversed such that the S source is brighter at 15 GHz. The different inter-band spectral index values between the two radio sources suggest unique AGN signatures. There are also several differences in absorption features between the two STIS UV/optical spectra, although some of these are artefacts due to uncleaned cosmic rays and noise and do not appear in the GMOS spectra. Finally, in the X-ray regime, the time-series of the two observations reveals the system to be dynamic, where the N source heavily dominates at first and then the S source flares, also leading to a flux ratio different from the other bands. \obj\ falls into the newly-discovered category of so-called nearly identical quasars \citep[$e.g.$,][]{Chan22}, where the properties of two closely separated quasars suggest lensing, but a foreground lensing galaxy is not apparent.  


Some of the apparent differences between the two quasars might be due to lensing-based flux anomalies caused by the foreground lensing galaxy. \citet{Xu12} find that dark matter halos and matter substructure ($e.g.$, from microlensing by stars) in simulated lensing galaxies can introduce a 20-30\% flux ratio anomaly for a lensed quasar at $z=2$. Observationally, evidence for the effects of foreground substructure has been been found using ALMA on the lensed system SDP.81 \citep{Hezaveh16}. \citet{Glikman23} have found flux anomalies between lensed images at different wavelengths. Based on simulations, they were not able to reconcile the observed wavelength-dependent flux anomalies with microlensing; however, their model of dark matter substructure in the lensing galaxy close to one imaged quasar was able to reproduce the high magnification needed to cause the flux anomaly in that wavelength. Since the apparent foreground galaxy in \obj\ is only marginally detected in NIR and 15 GHz radio and appears to be compact, it is difficult to say whether foreground substructure plays a part in the observed wavelength-dependent flux anomalies of the system. 

Radio interferometers at high angular resolution suffer from spatial brightness sensitivity filtering and thus introduce biases into component flux density measurements as compared to single dish observations, depending on the source substructure and scale of the interferometer. For poorly sampled $uv$ coverage, this can amount to a deficit in the observed flux densities. This could be contributing to the flux anomaly observed specifically in the higher-resolution Ku-band image.


Combining all of the evidence, we suggest that the most likely interpretation of \obj\ is that it is a single quasar which is gravitationally lensed into two images by a marginally detected foreground host galaxy, although this conclusion comes with the caveats noted above. In the remainder of the text, we probe the nature of the \obj\ system under this framework.
\subsection{Lensing Tests}\label{sec:lens_sim}
\begin{figure*}
     \centering

     \begin{subfigure}
         \centering
         \includegraphics[width=0.48\textwidth]{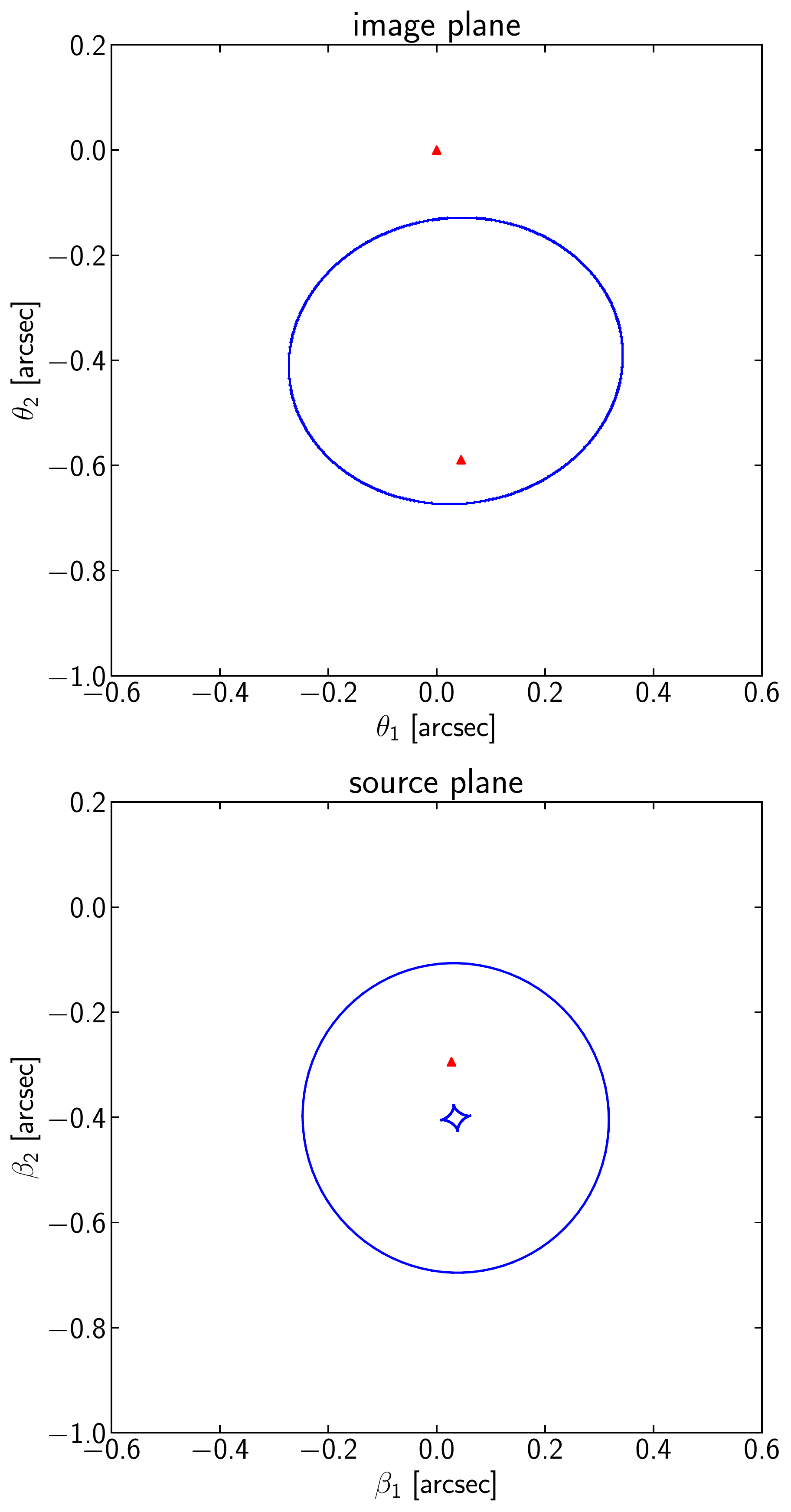}
     \end{subfigure}
     \hfill
     \begin{subfigure}
         \centering
         \includegraphics[width=0.48\textwidth]{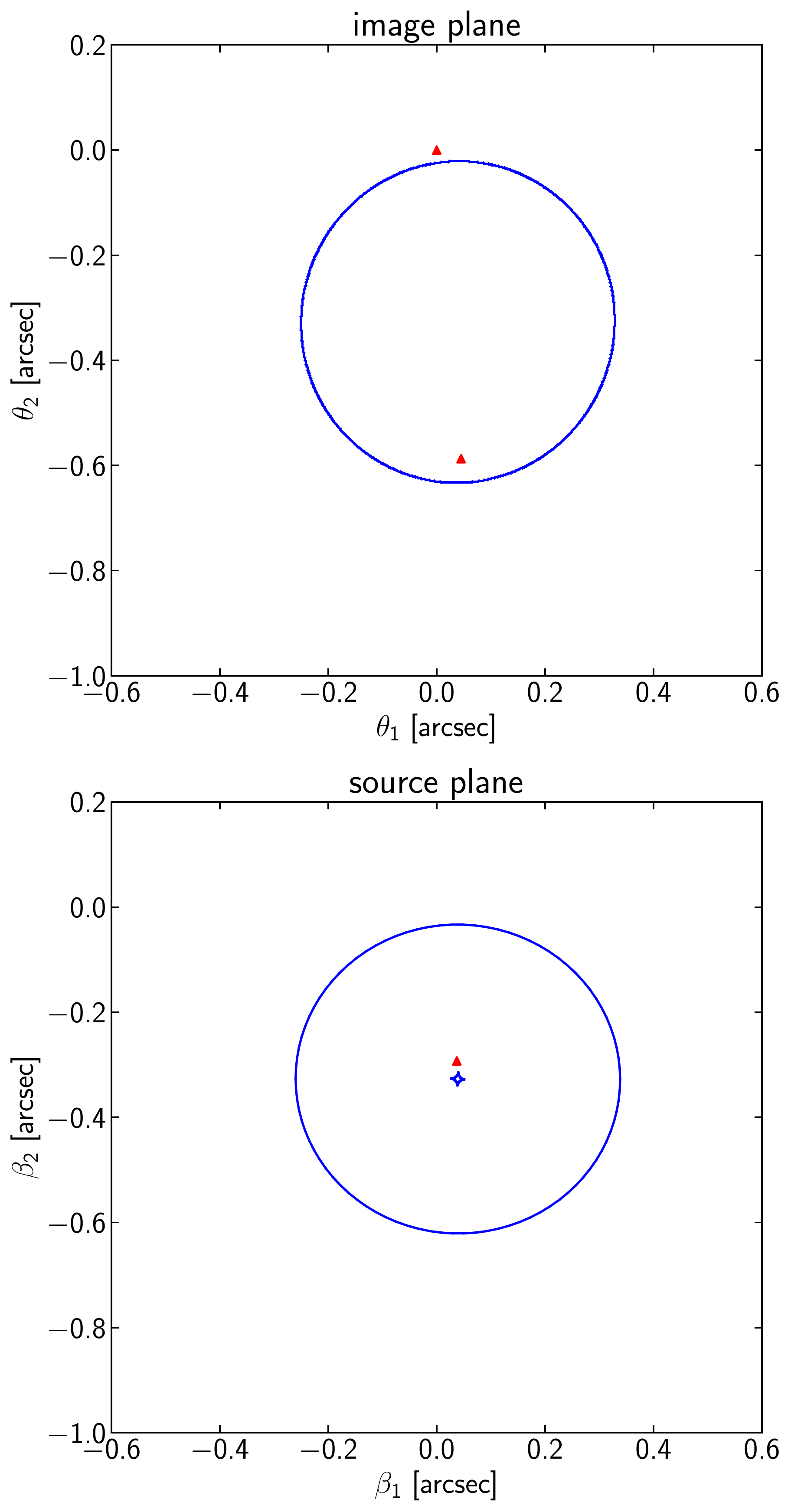}
     \end{subfigure}
        \caption{Lens Mass modeling for \obj. The columns give the decompositions for the HST F160W fitted results based on Model 1 (left) and Model 3 (right). {\bf Top:} image plane, where red triangles denote the observed relative positions of the N and S sources, the central blue dot is the location of the lensing galaxy as determined from our GALFIT analysis, and the blue ellipse is the critical curve. {\bf Bottom:} decomposed source plane using \texttt{glafic} \citep{Oguri10}, where the red triangle shows the position of the single quasar relative to the cut and caustic. Despite the differing moth models can be well-described by strong lensing using a simple SIE model.}
     \label{fig:lens}
\end{figure*}

We assess the feasibility of lensing to produce the observed source positions and magnitudes by conducting lens mass modeling using the software \texttt{glafic} \citep{Oguri10}. We test the effects of the differing flux ratios produced by the F160W Models 1 and 3. In Figure \ref{fig:lens}, we plot the de-projected source positions and the corresponding observed image positions. In both cases, the observed images are consistent with strong lensing. The positions of the image components relative to the lensing galaxy naturally explain the enhanced magnification observed for the N source in most of our observations. Given the on-sky separation of the images and the spectroscopic redshift of the quasar, we compute the minimum brightness needed for the foreground galaxy to produce the strong lensing using the fundamental plane relation of \citet{Faber76}, finding a minimum brightness of m$\sim$21.4 AB. Our best-fit values of m$_{\rm lens}\sim19.5-20.0$ AB are brighter than this threshold and thus are consistent with a foreground lensing galaxy at redshift of either $\sim0.3$ or $\sim1.6$.    

Given the rather different positions of each quasar in the image plane relative to the source quasar, we expect the different light travel paths to result in a lensing-based time lag between the S and N images. This is possibly corroborated by the X-ray observations which show variability over the course of $\sim$3 days between exposures. We obtain an estimate for the time lag using a relation for time delays in \texttt{glafic} based on a generalized isothermal potential \citep{Witt00,Oguri07, Lieu08}:
\begin{equation}
\Delta t_{ij}=\frac{1+z_{\rm l}}{2c}\frac{D_{\rm ol}D_{\rm os}}{D_{\rm ls}}(r^{2}_{j}-r^{2}_{i}),
\end{equation}
where the angular diameter distances $D$ are between the observer (o), lens galaxy (l), and source (s), and $r_{i}$ is the distance between image component $i$ and the center of the lensing galaxy. 
Assuming the lensing galaxy redshift of $z\sim1.6$, we find that the S source experiences a time delay of $\Delta t \sim 112$ (36) days behind the N source based on Model 1 (Model 3). If the lensing galaxy is actually at $z\sim0.3$, this time delay is much shorter, $\Delta t \sim 2.8$ (0.9) days. Clumps of matter along the line of sight and microlensing should not introduce any further time lags \citep{Lieu08}.

\begin{figure}
     \centering
         \includegraphics[width=0.48\textwidth]{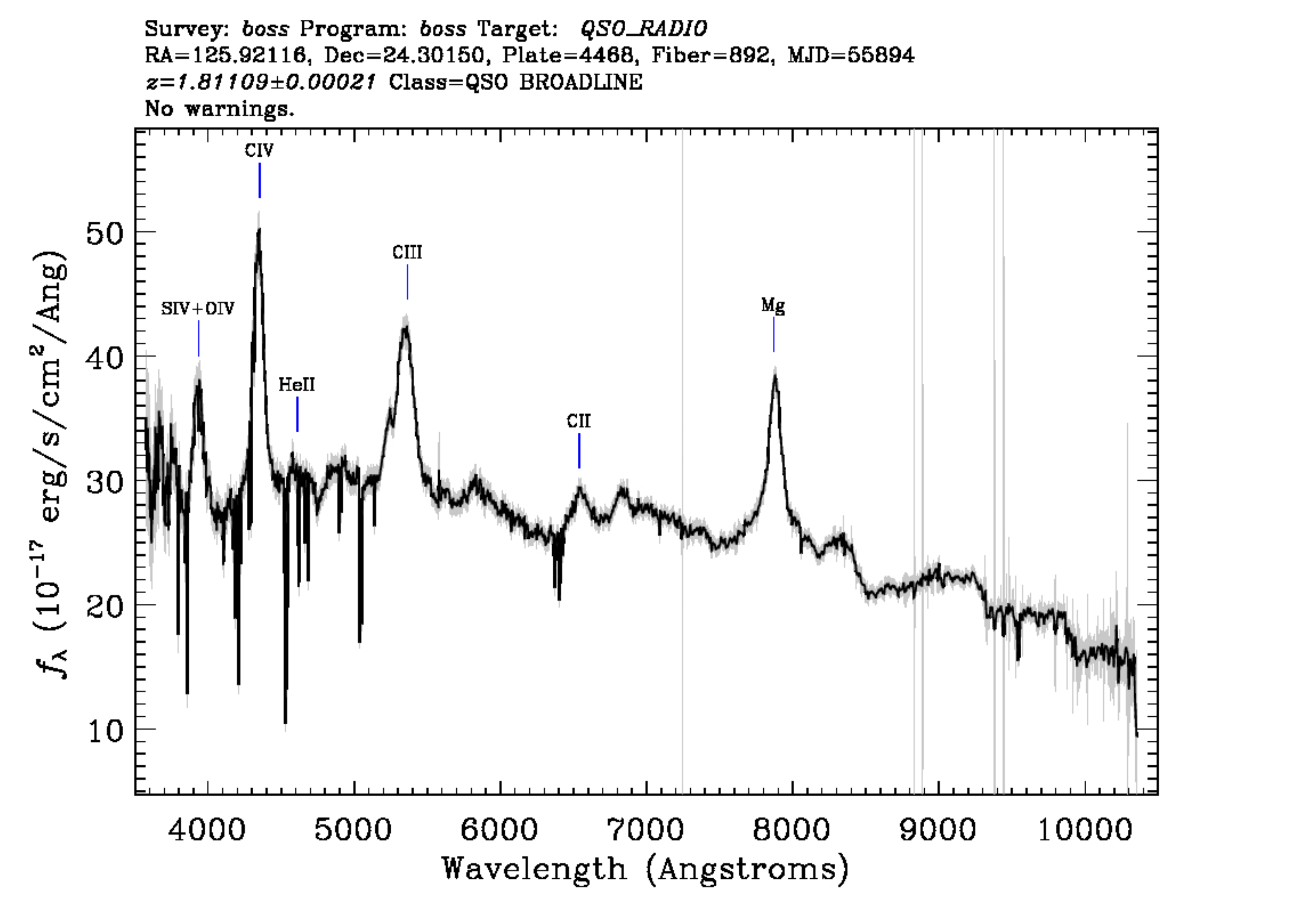}
        \caption{Archival SDSS spectrum of the \obj\ system, which has more wavelength coverage towards the observed-frame blue than the STIS or GMOS spectra. The presence of strong absorption lines around observed-frame 4500\AA\ and 4200\AA\ consistent with Mg and Fe suggest that an intervening absorber is located at $z\sim$0.62. These lines could be caused by relatively low column density gas in the intergalactic medium, but could be consistent with the foreground lensing galaxy. }
     \label{fig:boss}
\end{figure}


The archival SDSS spectrum for the system, shown in Figure \ref{fig:boss}, appears to exhibit absorption features that are not consistent with the redshift of the quasar system. Instead, the complexes at observed-frame $\sim4530$\AA\ and $\sim4200$\AA\ seem to be consistent with cold ISM absorption lines of \MgII\ 2796 and 2803$\lambda\lambda$ and \FeII\ 2586 and 2600$\lambda\lambda$, respectively \citep{Bowen00}. These absorption features suggest that the redshift of the absorber is $z\sim0.62$, which in turn implies time delays of $\Delta t \sim 7.5$ and 2.4 days for Models 1 and 3, respectively. Given the error on the best-fit magnitude of the lensing galaxy as well as the scatter of the Faber-Jackson relation, the lens galaxy redshift of 0.62 is not incompatible with the predictions of 0.3 and 1.6 from our lens mass modeling. At $z=0.62$, the lens galaxy would have an absolute magnitude of -22.88, similar to the elliptical galaxy M87. The Faber-Jackson relation predicts a minimum absolute AB magnitude of -21.73, which the observed lens galaxy magnitude satisfies.   

The \MgII\ and \FeII\ absorption features might not originate in the foreground lensing galaxy. There could be intervening intergalactic medium gas along the line of sight, and these low-ionization metal absorption features require only modest column densities to be produced \citep[$10^{16} <$ N(H{\sc i}) $\leq 10^{22}$ cm$^{-2}$,][]{Kacprzak11}. However, these \MgII\ absorption lines are known as a common marker in the spectra of lensed quasar systems. In a study of $\sim$266,000 SDSS quasars, \citet{Raghunathan16} found $\sim$37,000 systems across a range of redshifts with evidence of foreground \MgII\ absorption. As a rough estimate of how common these absorption features are in systems like \obj, we visually inspected $\sim$16 SDSS spectra of quasars at similar redshift and magnitude ($z\sim1.8$, $r$ mag$\sim17$), and found 4 quasars with strong absorption lines similar to those in the SDSS spectrum of \obj. We estimate the probability that a given system exhibiting \MgII\ absorption lines is a lensed quasar as:
\begin{align}
    &p({\rm lensing}\ |\ {\rm Mg II}) \\
    &= \frac{p({\rm Mg II}\ |\ {\rm lensing}) p({\rm lensing})}{p({\rm Mg II}\ |\ {\rm lensing}) p({\rm lensing}) + p({\rm Mg II}\ |\ {\rm dual}) p({\rm dual})} \\
    &\sim \frac{p({\rm lensing})}{p({\rm lensing}) + p({\rm Mg II}\ |\ {\rm dual}) p({\rm dual})} \\
    &= \frac{1}{1 + p({\rm Mg II}\ |\ {\rm dual}) p({\rm dual})/p({\rm lensing})} \\
    &\sim \frac{1}{1 + p({\rm Mg II}\ |\ {\rm dual}) }.
\end{align}
The first equation (lines 5 and 6) follows Bayes' theorem. The second equation assumes that $p({\rm Mg II}\ |\ {\rm lensing})=1$, $i.e.$ lensing systems always show \MgII\ absorption. The third equation divides both the numerator and the denominator by $p({\rm lensing})$ so one can see that the ratio of $p({\rm dual})/p({\rm lensing})$ also plays some role. The last equation assumes that $p({\rm dual})/p({\rm lensing})=1$, based explicitly on our results so far from two VODKA targets, but also informed by estimates from \citet{Chen22a} and \citet{Shen23} that the two quantities are likely of the same order of magnitude, resulting in our final expression. We lastly assume that $p({\rm Mg II}\ |\ {\rm dual}) = p({\rm Mg II}\ |\ {\rm normal\ quasars})$ since foreground absorption occurs randomly and the population of known dual quasars at similar redshift is too low for a statistical assessment. Of course, there are some uncertainties in the assumptions of $p({\rm Mg II}\ |\ {\rm lensing})=1$ (which may depend on S/N ratio and wavelength coverage) and $p({\rm dual})/p({\rm lensing})=1$, but it serves as a useful expression for us to estimate such probability. Having found $p({\rm Mg II}\ |\ {\rm normal\ quasars})\sim0.25$ above, we roughly estimate that \MgII\ absorption lines found in other VODKA targets have $\sim$80\% chance of indicating lensed quasars, making this a potentially powerful diagnostic test for ruling out dual quasar candidates when optical spectra are available. 

A time lag of roughly 2$-$3 days might explain the observed X-ray variability of the system. Given the $\sim$3 days between the two $Chandra$ observations, the N source is fairly luminous at first, perhaps exhibiting the onset of an X-ray enhancement. By the time of the second observation several days later, this signature has begun to fade by a factor of $\sim2.8$ in the N image location. Meanwhile, the S image has not yet seen the X-ray enhancement at the time of observation 1 and is thus undetected, but subsequently exhibits a brightening several days later to a luminosity comparable to the N image during observation 1, but $\sim2.6\times$ brighter than the N source during observation 2. The fact that $L_{\rm X}$ is slightly lower for S(obs 2) than N(obs 1) is likely a combination of the timing and the higher magnification for the N image. 

The X-ray variability observed for \obj\ is similar compared to other several other AGN. \citet{Niu23} have found X-ray flux variations in M81* on the scale of 2 - 5 days, and these flares were on the order of a factor of 2 brighter than the quiescent periods. M81* is smaller mass than the black hole in \obj\ by several orders of magnitude, and the X-ray variability of AGN is known to be anti-correlated with black hole mass \citep[][ and references therein]{Ludlam15}, suggesting a longer expected timescale. \citet{Huang23} found a $\sim5 \times$ increase in the X-ray flux of J081456.10+532533.5 between Chandra and XMM-Newton observations taken $\sim$30 days apart. During this time, they observed no noticeable variability in the optical, concluding that X-ray variability in AGN typically occurs on shorter timescales and at higher amplitudes than variability in the UV/optical range, as evidenced by a handful of archival type 1 quasars with strong X-ray flux variations and a lack of UV/optical variability \citep[][and references therein]{Huang23}. The STIS rest-frame UV spectra of \obj\ were taken $\sim$8 days apart from the X-ray observations, so it is plausible that the lack of UV deviations from the flux ratios observed in lower energy bands is not tied to the observed level of X-ray variability.



Lastly, we investigate further whether the faint blob observed in the VLA Ku-band can be explained by star formation radio emission from the lensing galaxy. Assuming all of the observed emission at 15 GHz is due to star formation, we extrapolate to the flux at 1.4 GHz assuming a power law with spectral index $\alpha = -0.8$. We convert the flux to a $k$-corrected luminosity density for each likely redshift \citep{Zakamska16}, as determined from the lens mass modeling and the SDSS spectrum. We then compute the expected star formation rate (SFR) due to nonthermal radio emission using the empirically-derived prescription of \citet{Bell03}.

We obtain values of SFR = 15.3, 81.8, and 770 $M_{\odot}$ yr$^{-1}$ at $z$ = 0.3, 0.62, and 1.6, respectively. Based on NIR samples from 3D-HST and the Cosmic Near-IR Deep Extragalactic Legacy Survey, typical star-forming galaxies at 1$\lesssim z\lesssim$2.5 have peak SFR $\sim10-15\ M_{\odot}$ yr$^{-1}$, which decreases to $\lesssim2\ M_{\odot}$ yr$^{-1}$ at $z=0$ \citep{VanDokkum13, Patel13}. If the blob is indeed the lensing galaxy and its radio emission is entirely due to star formation, then it is implausible that the galaxy resides at $z = 1.6$. While the SFR at $z = 0.62$ is also high, the actual rate could be lower if some of the observed radio emission is partially due to a weak AGN in the lensing galaxy. The lensing galaxy redshift of $\lesssim$ 0.62, and thus a time delay of $\sim$3 days, is thus not ruled out.

\subsection{Comparison to Other Lensed Quasars at High Z}\label{sec:compare}
Until very recently, the search for gravitational lenses had resulted in a only modest population of strongly lensed quasars. Some of the first well constrained examples were characterized using similar approaches of \hst\ image decomposition, yielding 8 systems at $z>1.5$ which have prominent foreground lensing galaxies \citep{Lehar00}. Large systematic searches for lensed quasars have been conducted in the optical, such as the SDSS Quasar Lens Search \citep[SQLS;][]{Oguri06b, Inada08}. Their sample of 11 lensed quasars contains 6 system at $z>1.5$, however they are all at source separations greater than 1\arcsec\ by design; the 10 additional lensed quasars that they studied are mostly at $z>2$, with several at smaller separations. Interestingly, SQLS also identified 30 genuine quasar pairs (sep$\lesssim$1\arcsec) of which 21 had been previously unrecognized. This unintended consequence highlights the complimentary nature of dual quasar and lensed quasar searches. 

More recently, the Gaia DR2 has been exploited to construct a catalog of 86 confirmed and 17 probable lensed quasars, some of which are at cosmic noon \citep{Lemon19, Lemon22}. However, the majority of these targets are also at larger separations than \obj\ (0.78$-$6.23\arcsec). The larger separations result from a component of the sample selection, where galaxy catalogs are matched to Gaia detections within a broad matching radius and the large PSF of the unWISE data. The authors find that, compared to mock catalogs, 56\% of the lensed quasar population at $z>1.5$ likely remains undiscovered and constitutes systems with small separations (sep $< 1.5$\arcsec) where the foreground lensing galaxy is faint and difficult to resolve from the quasar images \citep{Lemon22}. 

\citetalias{Chen22a} employed the varstrometry technique to identify promising candidates of dual quasars at smaller separations than those noted above. A related technique, the Gaia Multi Peak selection method \citep{Mannucci22, Mannucci23}, has recently uncovered several high-$z$ dual quasar candidates. Of the four candidates further investigated by \citet{Ciurlo23}, two systems had strong evidence for dual quasars based on spatially resolved H$\alpha$ emission from Keck OSIRIS. The J1608+2716 system shows three resolved nuclei at $z\sim2.6$, with separations between components under 0.3\arcsec. While one of the H$\alpha$ profiles of the components appears to be convincingly distinct, the remaining two line profiles have consistent line centers. \citet{Ciurlo23} suggest that the differing equivalent widths of the matching components could be due to microlensing effects, but despite the suggestive extended arc morphology of the system, a putative lensing foreground galaxy is not detected in their ground-based data. Similar spatially resolved spectroscopic follow-up could be used to distinguish lensed/dual quasars in the remaining VODKA sample.

Future campaigns using the Rubin Observatory Legacy Survey of Space and Time (LSST) will provide complimentary searches for lensed quasars. A mock catalog of simulated LSST images of lensed quasars by \citet{Yue22} demonstrates that the reasonable average seeing of the survey ($\sim0.7\arcsec$) will more than double the number of observable lensed systems with $z<7.5$ to roughly 2.4$\times10^{3}$ (although \citet{Lemon22} find that their mock catalog underpredicts the number of currently observed lenses by 44\%, so the final number of LSST-observable lenses might be higher). While most of their catalog contains quasars with image separations greater than that of \obj, smaller separation lenses can potentially be identified with other wide-area survey programs, such as with the Euclid Telescope, which will have a spatial resolution of 0.1 arcsec/pixel in optical wavelengths (0.3 arcsec/pixel for NIR imaging) across a 15,000 deg$^{2}$ area of sky \citep{Laureijs11}. The Nancy Grace Roman Space Telescope will also have spatial resolution comparable to HST (0.11 arcsec/pixel), but with $\sim200\times$ larger field of view and greater sensitivity (down to $\sim$27 AB mag) in NIR across a 2,000 deg$^{2}$ survey designed partially to target gravitational lensing \citep{Spergel15}. Identified candidates should then be followed up with multi-wavelength verification, similar to our approach above, to firmly separate out dual quasars from lensed systems.  

\section{Summary and Conclusions}\label{sec:sum}

In this work, we have focused on one of the most promising dual quasar candidates from the VODKA sample that is at a redshift near cosmic noon. To uncover the true nature of \obj, we rely on multi-wavelength imaging and spectroscopy. Based on the similarities between the flux ratios of the N and S quasars across several different bands, combined with the similarity of their spectra and the tentative detection of a central foreground galaxy, we conclude that the system is likely a single lensed quasar. This result is bolstered by our subsequent lens modeling, which shows that the best-fit positions and magnitudes from image decomposition are consistent with strong lensing. This assessment comes with several caveats due to conflicting results of our radio and X-ray imaging; however, these flux anomalies can potentially be explained by contributions from systematic effects, intrinsic variability, and lensing-based time delay of roughly 2$-$3 days. \obj\ is thus one of only $\sim$100 verified gravitationally lensed quasars at high-$z$; however, owing to our varstrometry selection technique, it is one of the smallest separation cases of these systems currently known.

As a case study, our analysis here gives us a road map for the rest of the VODKA candidates. When combined with the verification of SDSS J0749+2255 as a bona fide dual quasar \citepalias{Chen23}, we now have a series of tested metrics for robustly discerning between the dual versus lensed quasar scenarios. Specifically, we have found that several crucial pieces of multi-wavelength evidence are necessary for a conservative follow-up program. Spectroscopy is essential for probing the rest-frame optical or UV wavelengths where quasar emission lines are easily diagnosed while concurrently allowing for robust determinations of redshifts. This should be combined with imaging at optical or NIR wavelengths that can capture both faint distinguishing features, such as tidal tails, while also being able to spatially resolve potential foreground lensing galaxies. Our complimentary pairing of Keck and HST imaging in nearly identical bands has allowed us to capitalize on these two necessary, but often mutually exclusive advantages. Radio and X-ray imaging (and the corresponding general spectral features such as power law indices) can offer additional signs that two AGN are distinct. Based on the two VODKA targets studied in detail so far resulting in one dual quasar and one lensed, the VODKA selection method therefore appears to be just as likely to identify candidate lensed quasars at close separations as it is to find duals. Therefore, the dual population overall might be overestimated. Further multi-wavelength confirmation of candidates similar to this study is necessary to shed light on the true fraction of lensed versus dual quasars.


\acknowledgments

We thank A. Kemball for helpful discussions on strong gravitational lensing, as well as Junyao Li and Ming-Yang Zhuang for helpful discussions on PSF modeling and image decomposition. We also thank M. Leveille, A. Vick, R. Campbell, R. McGurk, J. Cortes, T. R. Geballe, S. Leggett, A. Nitta, T. Seccull, and H. Medlin for their help with our HST, Keck, Gemini, and VLA observations. HCH appreciates the discussion with Ting-Wen Lan on the foreground \MgII\ absorption.
This work is supported by NSF grant AST-2108162. YS acknowledges partial support from NSF grant AST-2009947. Support for Program 23700377 (PI: X. Liu) was provided by the National Aeronautics and Space Administration through Chandra Award Number GO2-23099X issued by the Chandra X-ray Center, which is operated by the Smithsonian Astrophysical Observatory for and on behalf of the National Aeronautics Space Administration under contract NAS8-03060. Support for Program number HST-GO-15900 (PI: H. Hwang), HST-GO-16210, and HST-GO-16892 (PI: X. Liu) was provided by NASA through grants from the Space Telescope Science Institute, which is operated by the Association of Universities for Research in Astronomy, Incorporated, under NASA contract NAS5-26555. This work was supported by JSPS KAKENHI Grant Numbers JP22H01260, JP20H05856, JP20H00181.

Based in part on observations obtained at the international Gemini Observatory (Program ID GN-2022A-Q-139; PI: X. Liu), a program of NSF’s NOIRLab, which is managed by the Association of Universities for Research in Astronomy (AURA) under a cooperative agreement with the National Science Foundation. on behalf of the Gemini Observatory partnership: the National Science Foundation (United States), National Research Council (Canada), Agencia Nacional de Investigaci\'{o}n y Desarrollo (Chile), Ministerio de Ciencia, Tecnolog\'{i}a e Innovaci\'{o}n (Argentina), Minist\'{e}rio da Ci\^{e}ncia, Tecnologia, Inova\c{c}\~{o}es e Comunica\c{c}\~{o}es (Brazil), and Korea Astronomy and Space Science Institute (Republic of Korea). This work was enabled by observations made from the Gemini North telescope, located within the Maunakea Science Reserve and adjacent to the summit of Maunakea. We are grateful for the privilege of observing the Universe from a place that is unique in both its astronomical quality and its cultural significance.

Based in part on data obtained at the W. M. Keck Observatory, which is operated as a scientific partnership among the California Institute of Technology, the University of California and the National Aeronautics and Space Administration. The Observatory was made possible by the generous financial support of the W. M. Keck Foundation. This research has made use of the Keck Observatory Archive (KOA), which is operated by the W. M. Keck Observatory and the NASA Exoplanet Science Institute (NExScI), under contract with the National Aeronautics and Space Administration. The authors wish to recognize and acknowledge the very significant cultural role and reverence that the summit of Maunakea has always had within the indigenous Hawaiian community. We are most fortunate to have the opportunity to conduct observations from this mountain.

This work is based in part on observations obtained with the 6.5 m Magellan-Baade telescope located at Las Campanas Observatory, Chile, and with the Apache Point Observatory 3.5 m telescope, which is owned and operated by the Astrophysical Research Consortium.

The National Radio Astronomy Observatory is a facility of the National Science Foundation operated under cooperative agreement by Associated Universities, Inc.

The scientific results reported in this article are based in part on observations made by the Chandra X-ray Observatory.

Based on observations made with the NASA/ESA Hubble Space Telescope, obtained from the Data Archive at the Space Telescope Science Institute, which is operated by the Association of Universities for Research in Astronomy, Inc., under NASA contract NAS 5-26555. These observations are associated with program GO-16210;270
(PI: X. Liu), and program GO-16892; PI:225
(X. Liu).

\newpage 
\bibliography{refs}
\end{CJK*}
\end{document}